\documentclass[aps, pra, reprint, groupedaddress, floatfix, longbibliography]{revtex4-1} 

\usepackage{graphicx}
\usepackage{epstopdf}
\usepackage{bm, amsmath, amssymb}
\usepackage{float}
\usepackage{placeins}
\usepackage{xcolor}
\usepackage{siunitx}
\usepackage[T1]{fontenc}
\usepackage{float}
\usepackage{enumitem}
\usepackage{hyperref}

\newcommand{\figwidth}{3.375in} 
\newcommand{\figwidthDouble}{7in}

\begin{document}

\title{Kerr-free three-wave mixing in superconducting quantum circuits}

\author{V. V. Sivak}
\email{vladimir.sivak@yale.edu}
\affiliation{Department of Applied Physics, Yale University, New Haven, CT 06520, USA}
\author{N. E. Frattini}
\affiliation{Department of Applied Physics, Yale University, New Haven, CT 06520, USA}
\author{V. R. Joshi}
\affiliation{Department of Applied Physics, Yale University, New Haven, CT 06520, USA}
\author{A. Lingenfelter}
\affiliation{Department of Applied Physics, Yale University, New Haven, CT 06520, USA}
\author{S. Shankar}
\affiliation{Department of Applied Physics, Yale University, New Haven, CT 06520, USA}
\author{M. H. Devoret}
\email{michel.devoret@yale.edu}
\affiliation{Department of Applied Physics, Yale University, New Haven, CT 06520, USA}

\date{\today}

\begin{abstract}

Quantum-limited Josephson parametric amplifiers are crucial components in circuit QED readout chains. The dynamic range of state-of-the-art parametric amplifiers is limited by signal-induced Stark shifts that detune the amplifier from its operating point. Using a Superconducting Nonlinear Asymmetric Inductive eLement (SNAIL) as an active component, we show the ability to {\it in situ} tune the device flux and pump to a dressed Kerr-free operating point, which provides a 10-fold increase in the number of photons that can be processed by our amplifier, compared to the nominal operating point. Our proposed and experimentally verified methodology of Kerr-free three-wave mixing can be extended to improve the dynamic range of other pumped operations in quantum superconducting circuits.

\end{abstract}

\maketitle

\section{Introduction}

Superconducting circuits offer the attractive possibility to synthesize systems with tailor-made Hamiltonians that display a variety of physical phenomena. Among the vast diversity of the Hamiltonians that can be produced, bilinear multimode bosonic Hamiltonians are the simplest, and yet can be wielded to generate nontrivial operations for quantum information processing. For example, single- and two-mode squeezing Hamiltonians are used to construct quantum-limited parametric amplifiers \cite{Clerk2008,Devoret2016}. Tunable-strength frequency conversion Hamiltonians enter in quantum state transfer between remotely separated modes \cite{Axline2018, Campagne-Ibarcq2018,Kurpiers2018}. Bilinear couplings with well-controlled phases are crucial for realizing active nonreciprocity in few-body systems such as parametric circulators and directional amplifiers \cite{Sliwa2015, Lecocq2017}, as well as for simulating many-body physics of topological band structure using photonic systems \cite{Fang2012} and implementing topological traveling-wave amplifiers \cite{Peano2016}. Furthermore, bilinear interactions are essential auxiliaries in nonlinear operations, such as the universal two-qubit entangling gate eSWAP \cite{Gao2018,Gao2018a} or Clifford gates on GKP-encoded logical qubits \cite{Gottesman2001,Terhal2016}. 

For such applications, it is important to master the implementation of bilinear Hamiltonians with high dynamic range. A common recipe for realizing a bilinear Hamiltonian is to pump the four-wave mixing nonlinearity of the Josephson junction with an external microwave drive. While this procedure synthesizes the target Hamiltonian, it unavoidably introduces additional spurious terms of higher order. To illustrate this inherent problem encountered in the task of Hamiltonian design in the realm of superconducting quantum circuits, consider the particular task of realizing the Hamiltonian of a degenerate parametric amplifier $H_{\rm DPA}=-\Delta_b b^\dagger b + g(b^2+b^{\dagger2})$, where $b$ refers to the photon annihilation operator in the rotating frame. It can be constructed with four-wave mixing by pumping $\overline{n}$ photons at detuning $\Delta$ into the Duffing oscillator realized with, for instance, a capacitively shunted Josephson junction or SQUID \cite{Eichler2013,Devoret2016}. In such an implementation, the junction four-wave mixing with strength $K$ induces three terms: (i)~the desired bilinear squeezing term with $g = K\overline{n}/2$, (ii) the spurious Stark shift $\Delta_b - \Delta = - 2K\overline{n}$, and (iii) the residual quartic term $\delta H = Kb^{\dagger 2}b^2/2$. Kerr terms of the type (iii) have in general serious consequences for the dynamic range of various pumped processes. For example, the photon-number-dependent rotation of the phase space caused by $\delta H$ leads to the distortion of quantum states of light in a microwave cavity at the level of a few photons \cite{Vlastakis2013, Kirchmair2013}, to ``bananization'' of squeezing \cite{Boutin2017,Malnou2018}, and to saturation of parametric amplifiers \cite{Eichler2013,Liu2017,Frattini2018,Planat2018}. In addition, the spurious Stark shift of the type (ii) limits the achievable gain in the multipumping schemes for directionality \cite{Sliwa2015, Lecocq2017} and deteriorates phase matching in traveling-wave amplifiers (TWPA) \cite{OBrien2014,Macklin2015,Vissers2015,Ranzani2018}.

Is it possible to generate pumped bilinear interactions without the detrimental sideeffects provided by Kerr? A necessary ingredient toward this goal is three-wave mixing, which is possible in superconducting circuits biased with external magnetic flux or DC current. For amplification, the third-order nonlinearity can be directly pumped to generate the squeezing term $g$ without the need for the Kerr term $K$. One example of such a three-wave mixing implementation of the DPA is based on flux-modulating the SQUID \cite{Yamamoto2008,Zhou2014,Simoen2015}. Alternatively, non-degenerate three-wave mixing is also available via the Josephson ring modulator (JRM) \cite{Abdo2013}, and is utilized in the Josephson parametric converter (JPC) \cite{Bergeal2010}. In both of these implementations the residual Kerr has been determined to cause amplifier saturation \cite{Zhou2014,Liu2017}. Recently a two-terminal three-wave mixing element, the Superconducting Nonlinear Asymmetric Inductive eLement (SNAIL), has been introduced \cite{Frattini2017,Frattini2018} as a tool to fight Kerr {\it in situ} by tuning the magnetic flux. In addition, three-wave mixing with Kerr suppression capability is also possible using RF SQUID \cite{Yurke1989, Zorin2016} and inductively shunted JRM \cite{Chien2019}.


However, as will be explained below, the Kerr constant $K$ can be significantly dressed by the presence of the pump. Therefore, the strategy to suppress Kerr must take into account this dressing. In this work, we demonstrate that it is indeed possible to realize the bilinear Hamiltonian corresponding to the degenerate parametric amplifier with three-wave mixing, while suppressing the effect of Kerr in the presence of the pump. We confirm the practicality of this Kerr cancellation by experimentally achieving an order of magnitude improvement in the saturation power and intermodulation distortion (IMD) properties of the parametric amplifier at a Kerr-free sweet spot, reaching the $1\rm\, dB$ compression power of $-102\rm\, dBm$ on par with the best published results \cite{Macklin2015}. We further show that this is possible without sacrificing nearly quantum-limited noise performance. More broadly, these results lead us to argue that the optimal strategy to engineer bilinear Hamiltonians should combine three ingredients: a) presence of a three-wave mixing capability, b) classical pump tone variable in strength and frequency and c) sufficiently versatile nonlinearity that allows for Kerr cancellation. 

\section{SNAIL Parametric Amplifier}

\begin{figure}[t]
 \includegraphics[width = \figwidth]{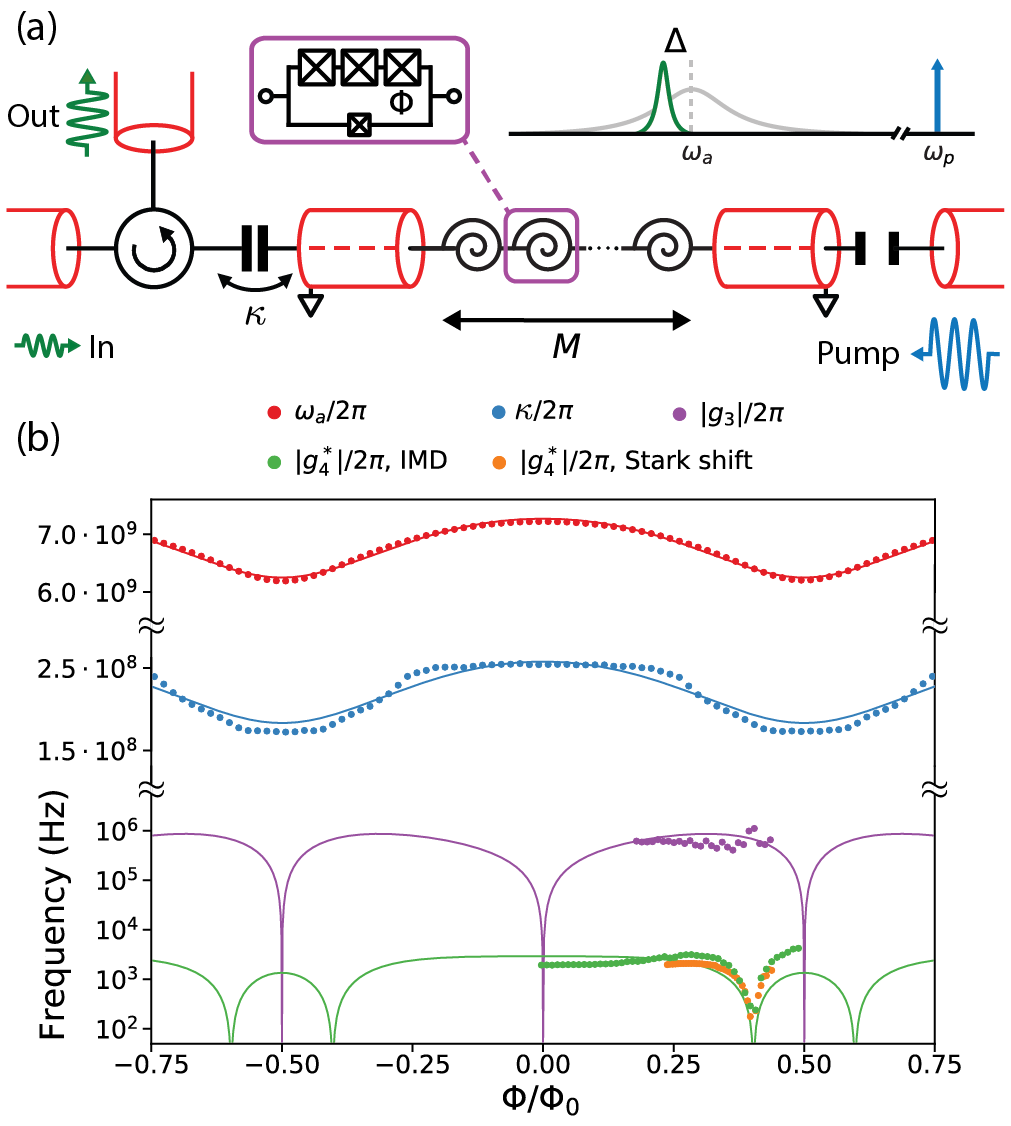}
 \caption{\label{fig1}  (a) Schematic of the reflection measurement setup. Insets show the circuit diagram of a single SNAIL and the frequency landscape of the parametric pumping process. (b)~Parameters of the SPA Hamiltonian and its dissipation as a function of magnetic flux, calculated (solid line) and measured (points) using the methods from \cite{Frattini2018}.}
\end{figure}

The device used in this work is the SNAIL parametric amplifier (SPA) \cite{Frattini2018}. Its electrical circuit, shown in Fig.~\ref{fig1}(a), consists of an array of $M=20$ SNAILs embedded into a microstrip transmission line resonator. The array is flux-biased, with a magnetic flux $\Phi$ piercing each loop. This knob enables the tuning of the SPA resonant frequency $\omega_a$ in the range $6.2-7.2\rm\, GHz$ and gives access to the cubic nonlinearity of the SNAIL's flux-dependent potential energy. The signal port of the device is strongly coupled to a $50\:\rm\Omega$ environment via a finger-capacitor and sets the energy damping rate of the SPA mode $\kappa/2\pi\in150-240\rm\, MHz$, while a weakly coupled pump port is used to supply the pump. The Hamiltonian of such system can be approximated as
\begin{align}
H_{\rm SPA}/\hbar= \omega_a(\Phi) a^\dagger a + \sum_{n=3}^\infty g_n(\Phi) (a^\dagger +a)^n.\label{H_spa}
\end{align}
The dependence of the most relevant Hamiltonian parameters on $\Phi$ is shown in Fig.~\ref{fig1}(b) and is calculated and measured using the methods elaborated in Ref.~\cite{Frattini2018} and in Appendices~\ref{supp1} and \ref{supp6}.

When a strong off-resonant pump is applied at frequency $\omega_p=2(\omega_a+\Delta)$, this leads to the effective Hamiltonian of a driven oscillator with Kerr nonlinearity
\begin{align}
H_{\rm eff}/\hbar=-\Delta_b b^\dagger b + g(b^2+b^{\dagger2}) + \frac{K}{2}b^{\dagger2}b^2,\label{b}
\end{align}
written in terms of the photon annihilation operator $b$ in the frame rotating at $\omega_p/2$. The parameters $\Delta_b$, $g$ and $K$ of the effective Hamiltonian in Eq.~\eqref{b} depend on all nonlinearities of the initial Hamiltonian in Eq.~\eqref{H_spa} and, importantly, also on the pumping condition. The lowest order dependence of these parameters on the average number of pump photons in the resonator $n_p$ is given by
\begin{align}
\Delta_b(n_p) & \approx \Delta - \left(\frac{32}{3}g_4 - 28\frac{g_3^2}{\omega_a}\right) n_p, \label{params1}\\
g(n_p) &\approx 2g_3\sqrt{n_p}, \label{params2}\\
K(n_p) &= 12g_4^* +O(n_p),\; {\rm where}\: g_4^*=g_4 - 5\frac{g_3^2}{\omega_a}.		\label{params3}
\end{align} 
These relations translate into the remarkable capability of the SPA to harness Kerr-free three-wave mixing: the parameters $g$ and $K$ are defined by different order nonlinearities, and their flux dependence is such that $K(\Phi)$ can go through zero and change sign while $g(\Phi)$ remains large enough to enable the operation of the SPA as an amplifier.

\begin{figure}[t]
 \includegraphics[width = \figwidth]{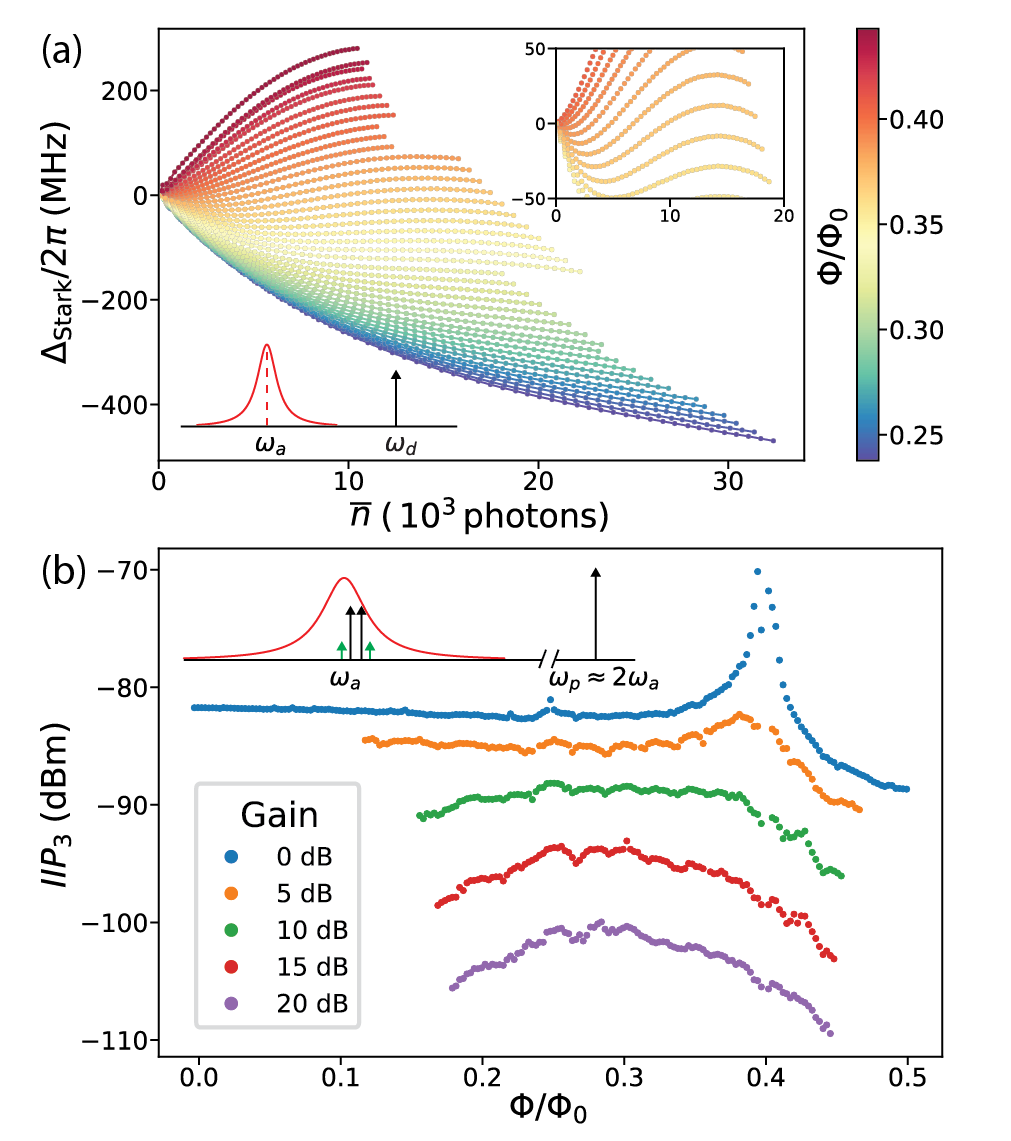}
 \caption{\label{fig2}  (a) Measured Stark shift $\Delta_{\rm Stark}$ versus the number of steady-state photons $\overline{n}$ in the resonator due to a drive at $\omega_d$, for different fluxes $\Phi$ denoted with color. The frequency landscape is sketched in the bottom left. The inset shows an enlargement of the region with suppressed Kerr. (b) $IIP_3$ measured with the configuration of tones sketched in black in the top left. The power of the pump at $\omega_p=2\omega_a$ is adjusted to produce different gains. The pump is turned off for $0\,\rm  dB$ gain.}
\end{figure}

The phase-preserving gain of such an amplifier at the signal frequency near $\omega_p/2$ in the presence of $n_s$ signal and $n_i$ idler intraresonator steady-state photons can be calculated in the input-output formalism using a semiclassical harmonic balance approximation as in Refs.~\cite{Zhou2014,Sundqvist2014,Frattini2018} and Appendix~\ref{supp4}. The result is given by 
\begin{align}
G = 1 + \frac{(2|g|\kappa)^2}{(\frac{\kappa^2}{4}+\Delta_i\Delta_s-4|g|^2)^2+\frac{\kappa^2}{4}(\Delta_i-\Delta_s)^2},
\label{gain}
\end{align}
where $\Delta_{s/i}(n_p,n_s,n_i)=\Delta_b(n_p)-K(n_p)\times(n_{s/i}+2n_{i/s})$. To obtain large gain, one needs to tune the model described by Eq.~\eqref{b} close to the parametric instability threshold at $n_p=(\kappa^2+4\Delta_b^2)/(8g_3)^2$. However, the spurious additional Stark shift $K(n_p)\times(n_{s/i}+2n_{i/s})$, created by the signal and idler photons, detunes the system from its operating condition, causing amplifier saturation. In order to improve the saturation power, we need to find an operating point at which $G$, given by Eq.~\eqref{gain}, is highly insensitive to $n_{s/i}$. To first order, this sensitivity is determined by $K$ dressed by the pump. As we will show next, in the presence of $10^3-10^4$ pump photons the $O(n_p)$ term in  Eq.~\eqref{params3} becomes comparable to the $n_p$-independent term, and the lowest-order perturbative approximation for $K(n_p)$ given by Eq.~\eqref{params3} breaks down. This breakdown is particularly important in the region of suppressed Kerr.

\section{Pump-induced dressing of Kerr}

Experimentally, we can acquire insight into the behavior of $K(n_p)$ at large numbers of steady-state pump photons by measuring the Stark shift and intermodulation distortion. Directly measuring the Stark shift coming from photons in the pump at $\omega_p$ is difficult in practice because it is obscured by the amplification process arising from the second term in Eq.~\eqref{b}. We instead measure the Stark shift caused by $\overline{n}$ steady-state photons of a strong near-resonant drive placed at $\omega_d=2\pi\times7.8\rm\, GHz$, about seven linewidths away from the resonance, which emulates the effect of the pump. The resulting $\Delta_{\rm Stark}(\overline{n})$ is shown in Fig.~\ref{fig2}(a) for different fluxes in the range $(0.24-0.44)\Phi_0$, where $\Phi_0=h/2e$ is the superconducting magnetic flux quantum. To the lowest order in $\overline{n}$, this Stark shift is given by $24g_4^*\overline{n}$. The upper bound on $\overline{n}$ in Fig.~\ref{fig2}(a) is roughly determined by the onset of chaotic behavior, caused by the excitation of free-particle-like states \cite{Verney2018}. The inset in Fig.~\ref{fig2}(a) shows that in a certain flux range we are able to suppress the linear contribution to the Stark shift and observe the beginning of an oscillation. We emphasize that this oscillation, similar to that theoretically predicted for a Josephson junction \cite{Kochetov2015,Verney2018}, occurs due to the higher-order terms in the SPA Hamiltonian given by Eq.~\eqref{H_spa}. This phenomenon corresponds to the dressing of the Kerr constant, which becomes especially relevant in the flux region of suppressed Kerr. 

We further confirm the pump-induced dressing of $K$ by investigating the IMD properties of the SPA. In this experiment, two near-resonant signals at $\omega_{s1}$ and $\omega_{s2}$ are sent to the resonator, and the sidebands at $2\omega_{s1}-\omega_{s2}$ and $2\omega_{s2}-\omega_{s1}$ appear at the output due to four-wave mixing. The input-referred third-order intercept point $IIP_3$ is then calculated as the extrapolated signal power for which the power in the sidebands is equal to the signal power. Here, it is related to the Kerr constant via
\begin{align}
IIP_3 = \frac{\kappa}{|K|}\frac{1}{(\sqrt{G}+1)^3}\hbar\omega_a \kappa, \label{IIP3}
\end{align}
as proved in \cite{Frattini2018}. The measurement result of $IIP_3$ with the pump off is shown in Fig.~\ref{fig2}(b) in blue, and the corresponding $g_4^*$ extracted according to Eqs.~\eqref{IIP3} and \eqref{params3} is shown in Fig.~\ref{fig1}(b) in good agreement with that extracted from the linear Stark shift. Eq.~\eqref{IIP3} suggests that $IIP_3$ decreases as $G^{-3/2}$ for large $G$ under the assumption that $K$ is independent of $n_p$ as in Eq.~\eqref{params3}, and, thus, the peak shape in $IIP_3$ at $\Phi=0.40\,\Phi_0$ would remain unchanged with increasing gain. However, in the experiment we find that pumping at $\omega_p=2\omega_a$ (i.e. $\Delta=0$) leads to the complete disappearance of the peak in $IIP_3$ at gains $G>5\rm\, dB$, meaning that $K(n_p(G))$ is no longer small. This dressing is most pronounced in the region of reduced Kerr where the contribution given by Eq.~\eqref{params3} is intentionally suppressed. Outside this region, the approximation $K(n_p)=\rm const$ works well up to $n_p$ required for $G=20\rm\, dB$, see Appendix~\ref{supp2}. 

We next formulate a procedure to correct for the dressing of $K$ by the pump. As suggested by the measurement of nonlinear Stark shift shown in Fig.~\ref{fig2}(a), changing $n_p$ at a fixed $\Phi$ can influence the sensitivity of the Kerr-induced terms $\Delta_{s/i}(n_p,n_s,n_i)$ in Eq.~\eqref{gain} to $n_{s/i}$, providing us the tool that we need in order to find the effective Kerr-free point. At a fixed $\Phi$, we can vary the pump photon number $n_p$ required to reach $G=20\rm\, dB$ by changing the pump detuning $\Delta$, which had been set to zero in the analysis so far. 

\section{Stability regions}

\begin{figure}[t]
 \includegraphics[width = \figwidth]{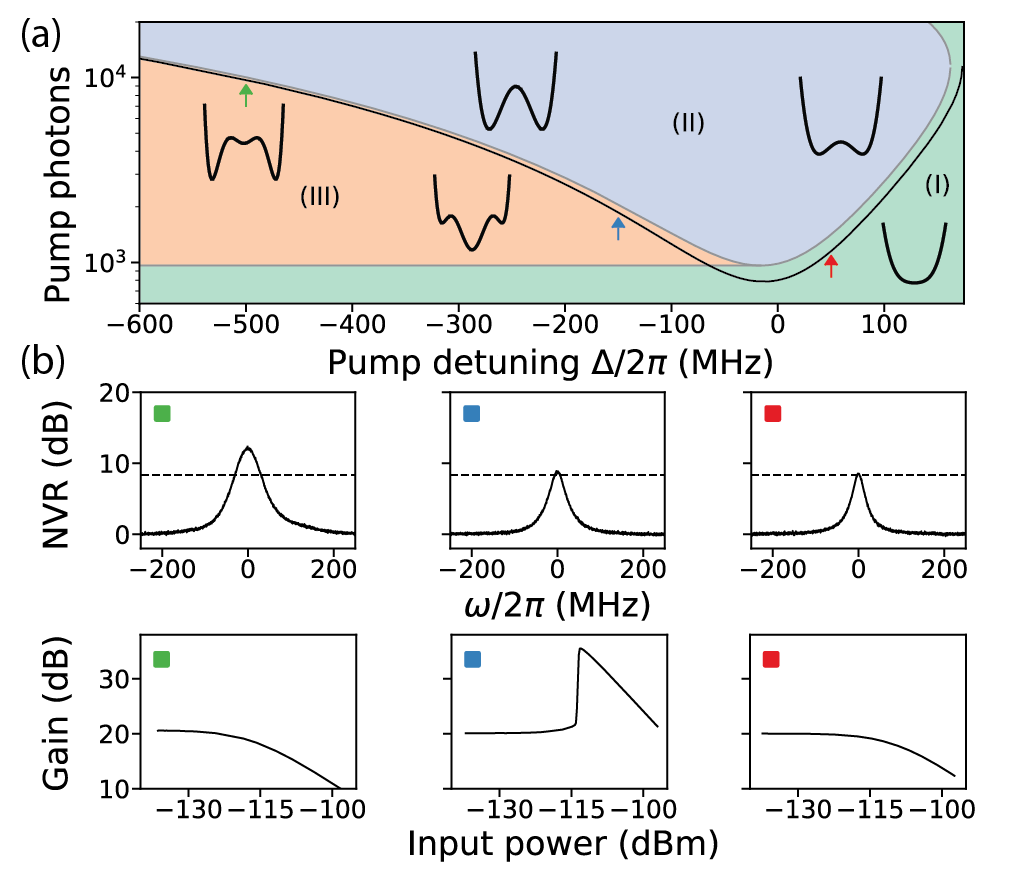}
\caption{\label{fig3} (a) Stability diagram of parametrically driven Kerr oscillator in coordinates $(\Delta,n_p)$, analytically calculated using approximations from Eqs.~\eqref{params1}-\eqref{params3} for the SPA parameters at $\Phi=0.30\,\Phi_0$, see Appendix~\ref{supp4}. Also shown are sketches of the effective 1D potential adapted from \cite{Zorin2011}. The black line below the parametric instability threshold is the line of $G=20\rm\, dB$. (b) Upper panels: noise visibility ratio at $G=20\rm\, dB$ as a function of noise frequency. The horizontal dashed line at 8.3~dB is a guide to the eye, indicating the quantum limit of the amplification chain. Lower panels: gain as a function of input signal power. The three data sets correspond to the operating points marked with arrows in (a).}
\end{figure}

The model described by Eq.~\eqref{b} exhibits a variety of new effects when $\Delta$ is varied away from zero. We first discuss these effects, which can be understood by considering the stability diagram of the Kerr oscillator driven parametrically near twice its resonant frequency, shown in Fig.~\ref{fig3}(a) for $\Phi = 0.30\Phi_0$. The colors in this diagram represent the three regions in the space of parameters $\Delta_b$ and $g$, in which the classical nonlinear dissipative dynamics of a complex amplitude $b$ has qualitatively different phase portraits \cite{Dykman1998,Wustmann2013}. In general, the motion of $b$ is not governed by a Hamiltonian that has separate potential and kinetic energy terms. However, under certain conditions, due to the separation of time scales one can integrate out the fast component of the motion and reduce the problem to the slow motion in an effective 1D potential \cite{Zorin2011,Lin2014}. This potential can have one, two or three minima depending on the parameters $\Delta_b$ and $g$, and is sketched for illustrative purposes in Fig.~\ref{fig3}(a). Such a 1D potential helps to qualitatively think about the problem, but the partition of the stability diagram into three separate single-, bi- and tristable regions can be done regardless, because it is based on the full 2D phase portrait of the system.


Region (I) of the stability diagram has a trivial ground state, in which $b$ fluctuates near zero. At $G=20\rm\, dB$ operating points in this region, the device behaves as a nearly quantum-limited amplifier which is seen from the measurement of the noise visibility ratio (NVR), defined as the ratio of noise power spectral densities with the SPA pump on and off, see Fig.~\ref{fig3}(b) right upper panel and Appendix~\ref{supp1}. In addition, the saturation curve taken at the signal frequency near $\omega_p/2$ monotonically decreases with increasing signal power, see Fig.~\ref{fig3}(b) right lower panel.

Region (II) has two stable points and, in the resulting ground state, $b$ has a finite mean amplitude, implying the resonator oscillates at frequency $\omega_p/2$ even with no applied signal power, an effect termed as ''parametric self-oscillation'' or ''period-doubling'' \cite{Dykman2012}. This region is separated from the rest of the diagram by the parametric instability threshold, where the gain of a parametric amplifier diverges \cite{Wustmann2013}. In the $K<0$ case, due to the Stark shift produced by $n_p$, there is a maximal positive detuning $\Delta$ beyond which large gain can no longer be obtained. Immediately next to the threshold, the black line denotes the pump photon number and detuning for $G = 20\rm\, dB$. In our experiment, we always operate on this line below the threshold and thus do not enter region (II),  although note that it is an interesting region for operating the device as a parametric oscillator \cite{Krantz2016,Lin2014}.

On the negative detuning side, the $G=20\rm\, dB$ line enters the region (III) of the oscillator stability diagram, which has three stable points. As long as the device operates at the single, global minimum at zero amplitude, it remains nearly quantum-limited as seen by the NVR trace in Fig.~\ref{fig3}(b) upper middle panel. Nevertheless, as shown in Fig.~\ref{fig3}(b) lower middle panel, the response of the amplifier to large signals changes qualitatively. When the signal power is increased above a certain threshold, the system undergoes the equivalent of a first-order phase transition and adopts a new working point with increased gain. This phenomenon is captured by the self-consistent equation \eqref{gain} for gain $G$ as a function of $n_{s/i}(G)$, as shown in Appendix~\ref{supp2}. The characteristic shape of the saturation curve in this regime was called ``shark fin'' in Ref.~\cite{Liu2017}, where a similar phenomenon was reported for the JPC. Interestingly, shark fins can be extremely sensitive to the level of input signal power, changing the output by $15\rm\, dB$ when the input difference is less than $1\rm\, dB$.

\begin{figure*}[t]
 \includegraphics[width = \figwidthDouble]{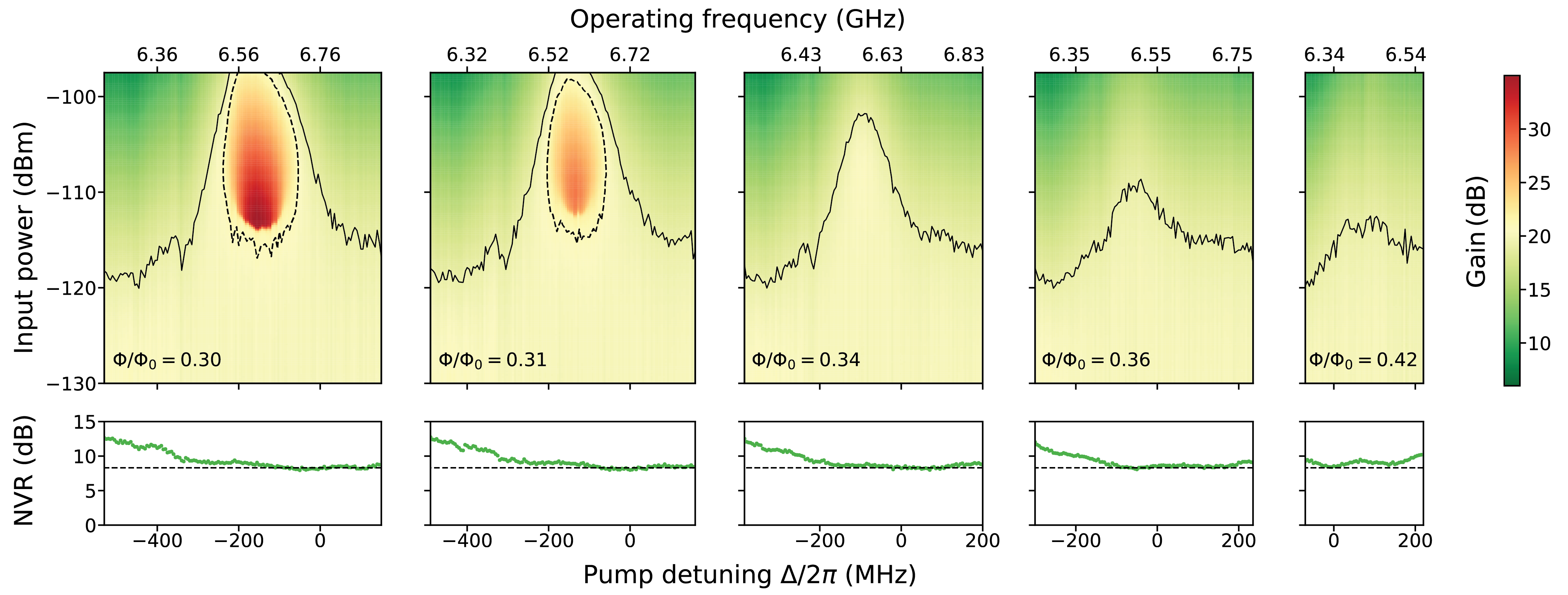}
 \caption{\label{fig4}  Upper panels: phase-preserving gain as a function of pump detuning $\Delta$ and signal power at various magnetic fluxes in the region of reduced Kerr. At each $\Delta$ the pump power is adjusted to reach $G=20\rm\, dB$ for signals weaker than $-140\rm\, dBm$. Solid and dashed black lines are isolines of $19\rm\, dB$ and $21\rm\, dB$ respectively. Lower panels: maximum noise visibility ratio as a function of pump detuning $\Delta$. The horizontal dashed line at $8.3\rm\, dB$ is a guide to the eye, indicating the quantum limit of the amplification chain.}
\end{figure*}

The presence of additional stable points in the region (III) implies that the period-doubling effect is also possible here. The associated instability where the system dynamically switches between the global minimum and the secondary ones  brings the danger of increased noise temperature of the amplifier. However, the high-amplitude minima are separated from the trivial minimum by potential barriers that prevent the system from switching \cite{Dykman1998,Svensson2017a}. Therefore, the NVR is not appreciably degraded at small red detuning inside region (III). Following the $20\rm\, dB$ curve in Fig.~\ref{fig3}(a), it is only beyond the detuning where all three coexisting steady states have equal  populations that the system prefers to stay in the high-amplitude state \cite{Lin2015}. Hence, we attribute the significant increase of NVR at large negative detuning, shown in Fig.~\ref{fig3}(b) upper left panel, to the enhanced switching between the high- and low-amplitude stable points. This effect is quantified by the increase of the noise temperature of the SPA by a factor of $2.9$ at $\Delta=-500\rm\, MHz$. In such a state, the amplifier is no longer quantum-limited and cannot be described by the conventional theory that leads to Eq.~\eqref{gain}. 

The stability diagram in Fig.~\ref{fig3}(a) was constructed at $\Phi=0.30\,\Phi_0$, away from the region of reduced Kerr. Since we cannot predict the stability regions at the fluxes where we expect the approximation in Eq.~\eqref{params3} to fail, we need to perform an empirical search for the dressed Kerr-free point in the space of parameters $\Phi$ and $\Delta$.

\section{Dressed Kerr-free point}

We show in Fig.~\ref{fig4} the results of the measurement of gain saturation (upper panels) and NVR (lower panels) for high-resolution sweeps of detuning $\Delta$ done at five different fluxes $\Phi$ in the region of reduced Kerr. At every detuning, the pump power is adjusted to obtain the small-signal gain of $20\rm\, dB$. The input signal power is then swept beyond the $1\rm\, dB$ compression point (solid black). The left and right boundaries of the $\Delta$ sweep at each $\Phi$ correspond to detunings where a small-signal gain $G=20\rm\, dB$ could not be achieved at any applied pump power. Moving from smaller to larger $\Phi$, the $\Delta$ tuning range decreases from about $900\rm\, MHz$ at $\Phi=0.19\,\Phi_0$ to about $200\rm\, MHz$ at $\Phi=0.48\,\Phi_0$ (not shown) and its median shifts from positive to negative values. 

We find the Kerr-free operating point in the presence of the pump to be at $\Phi=0.34\,\Phi_0,\,\Delta=-95\rm\, MHz$. There, the 1 dB compression power peaks at a value $P_{\rm 1 dB}=-102\rm\, dBm$ which is an order of magnitude larger than at $\Delta=0$. The tunable bandwidth over which we observe the $P_{1\rm dB}$ increase is about $200\; \rm MHz$. We measure a similar improvement in the IMD properties with peak $IIP_3=-94\rm\, dBm$, see Appendix~\ref{supp3}. Importantly, as indicated by the NVR, the amplifier remains nearly quantum-limited at this operating point. This result demonstrates that it is indeed possible to realize the degenerate parametric amplifier Hamiltonian with three-wave mixing, while suppressing the effect of Kerr in the presence of the pump.

\section{Conclusion}

To conclude, through theory and experiments we have formulated the methodology for realizing a quantum-limited degenerate parametric amplifier with suppressed spurious Kerr effect, and hence improved dynamic range. This work raises several directions for future investigation. Our device achieved $P_{\rm 1 dB}=-102\rm\, dB$, on par with the best published results, but with room for further improvement. In Appendix~\ref{supp5} we provide an outlook for how such an optimization would proceed and introduce a new concept of ``pump power efficiency'' that would need to be addressed to achieve further gains in dynamic range.

Another important direction would be to investigate the limits on the tunable ``Kerr-free bandwidth'' and on the compression power at the dressed Kerr-free point, which are possibly due to even higher-order terms not included in the effective Hamiltonian in Eq.~\eqref{b}. This naturally raises the question of whether it is possible to mitigate the detrimental effect of these higher order terms by circuit design and choice of pumping condition.

More broadly, we expect that the strategy of operating at a Kerr-free point while accounting for the dressing by the pump will be crucial for improved performance of other flux-biased three-wave mixing amplifiers that are, in principle, capable of operating at a Kerr-free point, such as the JPC with an inductively shunted JRM \cite{Chien2019} or the TWPA with array of RF SQUIDs \cite{Zorin2016} where Kerr cancellation will improve phase matching \cite{Zhang2017}. Finally, our methodology for achieving Kerr-free three-wave mixing can be applied for engineering other pumped bilinear Hamiltonians with superconducting circuits while suppressing the effect of spurious processes.

We acknowledge helpful discussions with S.~Mundhada, P.~D.~Kurilovich, V.~D.~Kurilovich, G.~Liu and T.~Roy. We also acknowledge the Yale Quantum Institute.  Facilities use was supported by the Yale SEAS clean room and YINQE. This research was supported by AFOSR under Grant No. FA9550-15-1-0029, and by ARO under Grants No. W911NF- 18-1-0212, W911NF-18-1-0020 and W911NF-16-1-0349.

\appendix

\section{Device parameters and measurement setup \label{supp1} }

\begin{figure*}[t]
 \includegraphics[width = \figwidthDouble]{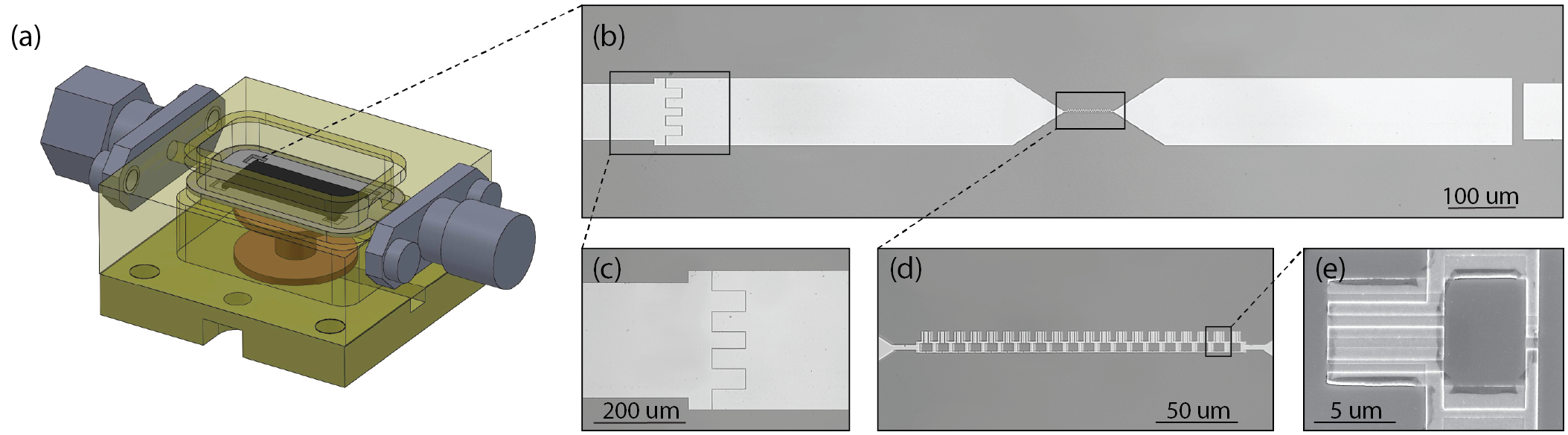}
 \caption{\label{S2} (a) Schematic of the SPA package that contains the superconducting box, SMA connectors, magnet spool, PCB and the silicon chip. (b) Optical microscope image of the microstrip resonator, (c) finger capacitor, (d) SNAIL array, and (e) electron micrograph of a single SNAIL.}
\end{figure*}

The SPA device package is shown in Fig.~\ref{S2}. Josephson junctions of the SNAIL are formed by $\rm Al/AlO_x/Al$ layers deposited using the Dolan bridge shadow evaporation process (critical current density $j_{c}=120\,{\rm A}/{\rm cm}^{2}$), and have critical currents $I_c=8.5\;\rm uA$ and $I_c=0.85\;\rm uA$. The microstrip transmission line resonator is formed by a $2\,\mu m$ thick silver film deposited on the back of a $300\,\rm \mu m$ thick silicon wafer to act as a ground plane, and by the aluminum traces on the chip. The silver back plane of the chip is glued using conducting silver paste to the copper back plane of the ${\rm TMM}10{\rm i}$ printed circuit board (PCB), which is soldered to the gold-plated aluminum box, see Fig.~\ref{S2}(a). The first resonant mode of the box is at $18\,{\rm GHz}$, well above relevant frequencies for the operation of the SPA. The pump and signal aluminum transmission line traces on the chip are wire bonded to the copper transmission line traces on the PCB, which are soldered to the edge-mount SMA connectors. 

A superconducting NbTi coil that applies magnetic flux to the SNAILs is mounted under the PCB inside of the box, as shown in Fig.~\ref{S2}(a). The aluminum box acts as a partial shield against stray external magnetic fields. For better protection, an additional cryogenic $\mu$-metal shield is fitted around the aluminum box. The device is thermally anchored to the base stage of a dilution refrigerator ($T=24\rm\, mK$) through this shield. It is mounted back-to-back with the $4-8{\rm \,GHz}$ cryogenic circulator, which is connected to the signal port of the SPA with a short 2 inch cable. This configuration allowed us to significantly suppress the small-scale ripples in the data resulting from the impedance mismatches \cite{Mutus2014,Roy2015,Frattini2018}. We found that the close proximity of the commercial ferrite circulator does not influence the SPA performance.

The SPA pump tone is generated by an Agilent PSG E8257D generator. The DC current is supplied to the coil by a Yokogawa GS200 current source. All measurements, including the IMD and NVR, are done using the corresponding measurement classes of a Keysight PNA-X N5242A network analyzer. 

We found that PNA-X ``Noise Figure'' measurement class sometimes displays a sharp spike at $\omega_p/2$, which is absent in a separate measurement done with an Agilent EXA N9010A Spectrum Analyzer, and likely results from the lower tolerance of the PNA-X to the pump leaking through the isolators of the output chain at frequencies close to their cutoff at $12\,\rm GHz$. This spike does not affect the rest of the data, as verified by the independent measurements of NVR, and is filtered out in software. 

Extraction of the amplifier noise temperature from NVR requires the calibration of the gain of the output chain, for example with the shot noise thermometer \cite{Spietz2003,Spietz2010}, which was not available in our setup. However, our claim that the system is nearly quantum-limited when ${\rm NVR}\approx 8.3\rm\, dB$ is supported by the independent measurement of quantum efficiency $\eta=0.6$ in the experiment \cite{Touzard2018} done with the SPA.

\section{Intermodulation distortion (IMD)  \label{supp3} }

\begin{figure}[b]
 \includegraphics[width = \figwidth]{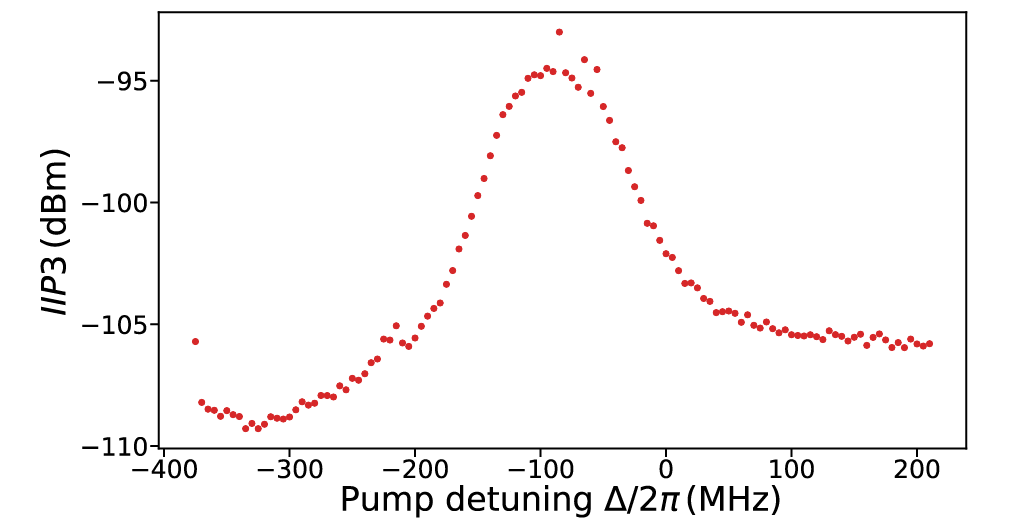}
 \caption{\label{S3} $IIP_3$ as a function of pump detuning measured in the same sweep as the data in Fig.~\ref{fig4} at $\Phi=0.34\,\Phi_0$.}
\end{figure}

In the IMD measurement the two signals at $\omega_{s1}$ and $\omega_{s2}$ of equal power are sent to the input of the SPA. The median of the signals is detuned from the center of the Lorenzian gain at $\omega_p/2$ by amount $\delta_1\equiv(\omega_{s1}+\omega_{s2})/2 - \omega_p/2=2\pi\times500\rm\, kHz$ to avoid phase-sensitive amplification. The detuning is chosen small compared to the amplifier 3dB-bandwidth $B\sim2\pi\times 25\rm\, MHz$. The separation of the two tones is $\delta_2=\omega_{s2}-\omega_{s1}=2\pi\times100\rm\, kHz$. Due to the nonlinear intermodulation distortion, the sidebands at $2\omega_{s2}-\omega_{s1}$ and  $2\omega_{s1}-\omega_{s2}$ appear at the output and are measured using the IMD measurement class of the PNA-X. The $IIP_3$ extracted from the raw data is shown in Fig.~\ref{S3}, and exhibits a peak value $IIP_3=-94\rm\, dB$ at the dressed  Kerr-free point similar to the peak in $P_{1\rm dB}$.

\section{Saturation power and shark fins\label{supp2} }

Saturation of a good amplifier happens at large average signal and idler photon numbers $n_{s/i}\gg1/2$, justifying the use of the semiclassical approximation, in which the phase preserving gain $G$ at the signal frequency near $\omega_p/2$ is given by Eq.~\eqref{gain}. The small-signal limit of this expression is given by
\begin{align}
G_0=1+\frac{4\kappa^2|g|^2}{( \Delta_b^2+\frac{\kappa^2}{4} -4|g|^2)^2}. \label{S_G0}
\end{align}
The goal of this section is to derive a closed-form expression for the input signal power $P_{\rm in}$ at which the gain changes from $G_0$ to $G$ (which directly gives the closed-form expression for $P_{1\rm\,dB}$) and to study the phenomenon of shark fins. Using input-output theory, we can establish the relation between the input power $P_{{\rm in}}$ and the intraresonator populations $n_{s}$ and $n_{i}$. For large gain $G\gg1$ we obtain $n_{s}\approx n_{i}\approx GP_{{\rm in}}/\hbar\omega_{a}\kappa$ with a relative error $\delta n_s/n_s\sim1/\sqrt{G}$, which is about $10\%$ for $G=20\,\rm dB$. In this approximation, we can rewrite Eq.~\eqref{gain} in the form
\begin{align}
G=1+\frac{4\kappa^{2}|g|^{2}}{(\Delta_{{\rm eff}}^{2}+\frac{\kappa^{2}}{4}-4|g|^{2})^{2}},\\ \label{eq:S_gain}
\Delta_{\rm eff}=\Delta_{b}-3K\frac{GP_{{\rm in}}}{\hbar\omega_{a}\kappa}, 
\end{align}
where $\Delta_{\rm eff}$ contains the additional Stark shift created by the signal and idler photons (this formula is given in \cite{Frattini2018}, but with a typo which is corrected here).  After plugging this into Eq.~\eqref{eq:S_gain} and solving for $P_{{\rm in}}$,
we obtain
\begin{align}
P_{{\rm in}}=\frac{\hbar\omega_{a}\kappa}{3KG}\bigg(\Delta_{b}\pm\sqrt{4|g|^{2}-\frac{\kappa^{2}}{4}+\frac{2\kappa|g|}{\sqrt{G}}}\bigg).
\end{align}

\begin{figure*}[t]
 \includegraphics[width = \figwidthDouble]{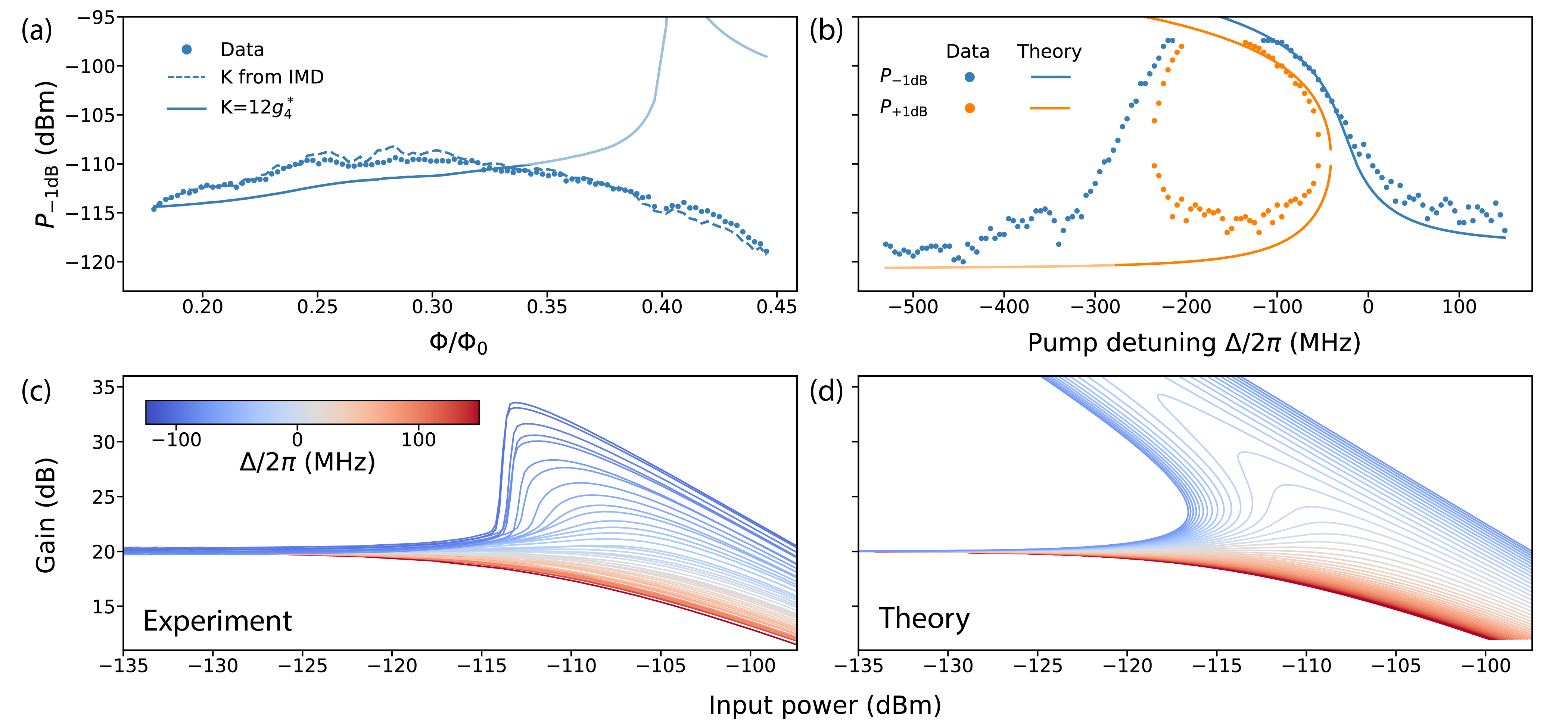}
 \caption{\label{S1} Input saturation power (a) as a function of $\Phi$ at fixed detuning $\Delta=0$, and (b) as a function of $\Delta$ at fixed flux $\Phi=0.30\,\Phi_0$. Expression \eqref{eq:S_P1dB}  has limited validity, it is plotted at $\Phi\gtrsim0.30\,\Phi_0$ only to show the expected divergence, but should not be extrapolated to the region of significant Kerr dressing. Empirically, the region of validity corresponds to where the blue-dashed and solid theory lines in (a) are similar. Likewise, it should not be extrapolated beyond the detuning $\Delta\approx-300\rm \, MHz$ in (b), where the ground state of the metapotential in Eq.~\eqref{b} is significantly modified and the derivation of Eq.~\eqref{eq:S_P1dB} is not valid. Panels (c) and (d) show the experimental and theoretical saturation curves for various pump detunings $\Delta$ denoted with color. At negative $\Delta$ the system undergoes an equivalent of the first-order phase transition with increased input signal power. }
\end{figure*}

The Hamiltonian parameter $g$ can be expressed through $G_{0}$ from Eq.~\eqref{S_G0}, which leads to
\begin{align}
P_{{\rm in}}=\frac{\hbar\omega_{a}\kappa^{2}}{3KG}\bigg\{\frac{\Delta_{b}}{\kappa}\pm\sqrt{\frac{\Delta_{b}^{2}}{\kappa^{2}}+\frac{\sqrt{G_0}-\sqrt{G}}{\sqrt{G_0G}}\sqrt{\frac{\Delta_{b}^{2}}{\kappa^{2}}+\frac{1}{4}}}\bigg\}. \label{bbb}
\end{align} 

In this expression both $K$ and $\Delta_{b}$ depend on $n_{p}$, and thus implicitly on gain. When we consider a simplified case of small negative $K$, such that $\Delta_{b}\approx\Delta$ and $K\approx 12g_4^*$, we can further simplify Eq.~\eqref{bbb} to
\begin{align}
P_{{\rm in}}=\frac{\hbar\omega_{a}\kappa^{2}}{36g_{4}^*G}\bigg\{\frac{\Delta}{\kappa}\pm\sqrt{\frac{\Delta^{2}}{\kappa^{2}}+\frac{\sqrt{G_0}-\sqrt{G}}{\sqrt{G_0G}}\sqrt{\frac{\Delta^{2}}{\kappa^{2}}+\frac{1}{4}}}\bigg\}.\label{eq:S_P1dB}
\end{align}

At positive detuning, only one solution is possible for $G<G_{0}$ and no solutions for $G>G_{0}$, while at negative detuning two solutions are possible for $G>G_{0}$ and one for $G<G_{0}$, which corresponds to the ``shark fin'' shape of the saturation curve.

The comparison of Eq.~\eqref{eq:S_P1dB} to the measured  $P_{1{\rm dB}}$ at a fixed detuning $\Delta=0$ as a function of flux $\Phi$ is shown in Fig.~\ref{S1}(a), and comparison to  the measured $P_{1{\rm dB}}$ at a fixed flux $\Phi=0.30\,\Phi_0$ as a function of detuning $\Delta$ is shown in Fig.~\ref{S1}(b). For the plotted theory lines, we have assumed that there exists a systematic miscalibration of the resonant frequency (and therefore the detuning $\Delta$) by approximately $25\rm\, MHz$. The compression power is very sensitive to such miscalibration especially near the detuning $\Delta=0$,  as evident from Fig.~\ref{S1}(b). Accounting for this shift aligns the theory and data much better in both $\Delta$ and $\Phi$ axes, while without this correction we find a $6\rm\, dB$ systematic disagreement between theory and data in Fig.~\ref{S1}(a). 

The extrapolation of Eq.~\eqref{eq:S_P1dB} to detunings $\Delta<-300\rm\, MHz$ predicts persisting shark fins, which are not observed in data. We understand this effect to arise from the fact that the ground state of the metapotential in Eq.~\eqref{b} is significantly modified at large negative detuning $\Delta$  due to the secondary high-amplitude minima, which invalidates the derivation of $P_{1\rm dB}$ presented in this section. Empirically, we locate this region by the increased NVR at large negative detuning, as discussed in the main text. Similarly, the extrapolation of Eq.~\eqref{eq:S_P1dB} to fluxes $\Phi\gtrsim~0.30\Phi_0$ leads to the incorrect prediction of improved compression power at an unpumped Kerr-free point at $\Phi=0.40\,\Phi_0$, as shown in Fig.~\ref{S1}(a). Using Eq.~\eqref{eq:S_P1dB} with $K$ extracted from the IMD measurement at gain $G=20\rm\, dB$ leads to a much better functional agreement, indicating that the IMD experiment better captures the dressing of the Kerr constant by the pump.

We can view Eq.~\eqref{eq:S_gain} as a self-consistent equation of state, analogous to the Van der Waals equation for nonideal gas, which exhibits a first-order phase transition \cite{Landau}. In this analogy, the detuning $\Delta$ plays the role of temperature, input signal power $P_{\rm in}$ plays the role of pressure, and gain $G$ plays the role of volume. As shown in Fig.~\ref{S1}(d), for the negative detuning $\Delta$, the equation \eqref{eq:S_gain} predicts the possible existence of a metastable state with high gain. In the sweep of input signal power, the system switches to this state along the line analogous to the isobar in the Maxwell construction, as shown in the experimental data in Fig.~\ref{S1}(c).

Finally, note that the flux tuning range of the SPA shown in Fig.~\ref{S1}(a) corresponds to a wide tunable bandwidth of about $1\rm \, GHz$. Moreover, at a fixed $\Phi$ the tunability can be accomplished with the help of pump detuning $\Delta$, which corresponds to a tuning range of approximately $700\rm\, MHz$, as shown in Fig.\ref{S1}(b) for $\Phi=0.30\,\Phi_0$. However, not all of this range corresponds to the quantum-limited performance as discussed in the main text, reducing the usable tunability at a fixed flux to about $400\rm\, MHz$. Due to this property, such an SPA is a good candidate for time-multiplexed readout schemes, in which the pump frequency can switch to match the required readout channel, similarly to the proposal of flux switching in Ref.~\cite{Abdo2017}.

\section{Stability diagram of the parametrically driven Kerr oscillator  \label{supp4} }

In this section we derive the separation lines between the regions of the stability diagram at $\Phi=0.30\,\Phi_0$ shown in Fig.~\ref{fig3}(a), using the semiclassical harmonic balance method. When the pump and signal tones are incident on the SPA nonlinear resonator, the mixing between them results in the amplification of the signal, creation of the idler and creation of other intermodulation products at frequencies that can be parametrized as $\omega_{nm} = n\omega_p + m\omega_s$ where $n$ and $m$ are integers. However, under certain conditions, generation of radiation which does not belong to this family of intermodulation products is possible. Of particular interest to us is the coherent period-doubling tone at frequency $\omega_p/2$ generated solely by the pump.

To grasp the major processes that happen in the pumped SPA, we perform a second-order harmonic balance calculation which accounts for the following harmonics:
\begin{itemize}[leftmargin=*]
\item[$\circ$] Pump $\omega_p$, signal $\omega_s$ and idler $\omega_i=\omega_p-\omega_s$ comprise a minimal set required for quantum-limited amplification.
\item[$\circ$] $2\omega_p$, $2\omega_s$, $2\omega_i$, $\omega_p+\omega_s$, $\omega_p+\omega_i$, $\omega_s-\omega_i$ and $\rm DC$ are required in the consistent calculation that takes into account linear Stark shift. These harmonics lead to $\sim g_3^2/\omega_a$ corrections to $K$ in Eq.~\eqref{params3} and to pump-induced Stark shift in Eq.~\eqref{params1}. 
\item[$\circ$] $\omega_p/2$, $\omega_p/2-\omega_s$ and $\omega_p/2-\omega_i$ account for a possible period-doubling effect.
\end{itemize}

\begin{figure*}[t]
 \includegraphics[width = \figwidthDouble]{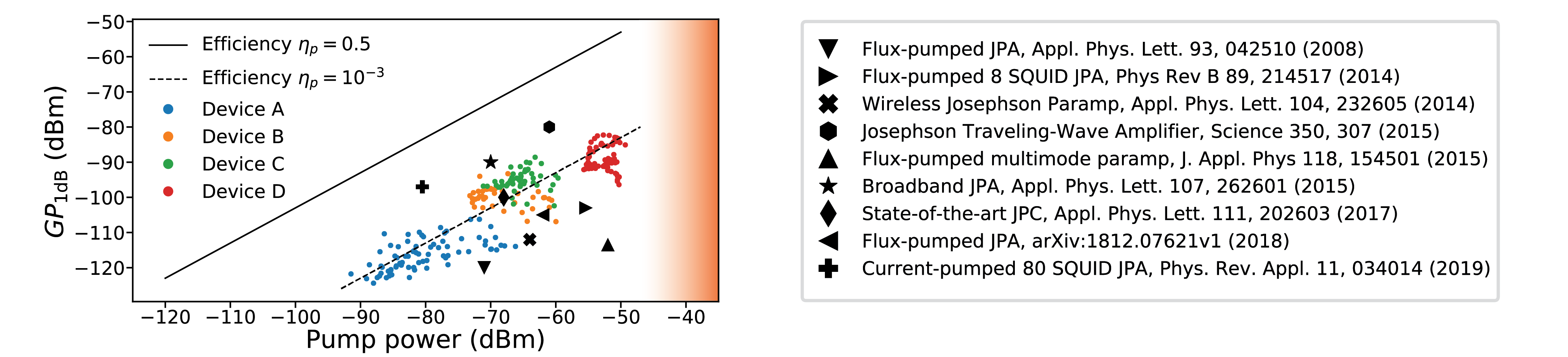}
 \caption{\label{S4} Power efficiency at $G=20\,\rm dB$ of various amplifiers available in the literature. Data for SPAs (devices A-D from Ref.~\cite{Frattini2018}) corresponds to operating points at different fluxes. The line of $\eta_p=0.5$ is shown for reference, and is not achievable even in theory; this value is chosen to indicate equal pump power distribution between signal and idler. The color gradient on the right represents the possible increase of fridge base temperature at high pump powers.}
\end{figure*}

Following the steps of \cite{Frattini2018} and Refs.~\cite{Zhou2014,Sundqvist2014}, we derive the self-consistent system of equations that links the amplitudes of all chosen harmonics. This is done by solving the quantum Langevin equation (QLE) \cite{Gardiner2004} without RWA approximation (because the out-of-band harmonics do not satisfy it). In order to do this, we need to take into account that the negative frequency Fourier component of the annihilation operator $a$ is linked to the positive frequency component via the relation
\begin{align}
a_{-\omega}^\dagger = \frac{\omega_a-\omega}{\omega_a+\omega}a_\omega. \label{neat_formula}
\end{align}
The $a_{-\omega}^\dagger$ is usually dropped from the frequency-domain QLE for $a_\omega$ if all the signals arriving at the resonator are near the resonance. Eq.~\eqref{neat_formula} can be derived from the relation $\dot{\Phi}=C_a Q$ between the canonical coordinate $\Phi$ of the mode and its canonical momentum $Q$, written in second-quantized form.

Taking into account Eq.~\eqref{neat_formula}, we seek the semiclassical harmonic balance solution to QLE in the form 
\begin{align}
\alpha(t) = \sum_x \left(\alpha_x e^{-i\omega_x t}+\frac{\omega_a-\omega_x}{\omega_a+\omega_x}\alpha_x^* e^{i\omega_x t}\right),
\end{align}
where $x$ runs over all harmonics described previously.

Equations for the out-of-band harmonics  can be partially solved, and the system further reduced to three complex equations for the amplitudes at $\omega_s$, $\omega_i$ and $\omega_p/2$, denoted as $\alpha_s$, $\alpha_i$ and $\alpha_h$, respectively, as follows:
\begin{widetext}
\begin{align}
(\omega +\Delta_b+i\kappa/2)\alpha_s & = u_s + (4g_3\alpha_p+12g_4^*\alpha_h^2)\alpha_i^* + 12g_4^*(|\alpha_s|^2+2|\alpha_i|^2+2|\alpha_h|^2)\alpha_s,\label{x1}\\
(-\omega+\Delta_b+i\kappa/2)\alpha_i & = u_i + (4g_3\alpha_p+12g_4^*\alpha_h^2)\alpha_s^* + 12g_4^*(|\alpha_i|^2+2|\alpha_s|^2+2|\alpha_h|^2)\alpha_i,\label{x2}\\
(\Delta_b+i\kappa/2)\alpha_h & = u_h + 4g_3\alpha_p\alpha_h^* + 12g_4^*(|\alpha_h|^2+2|\alpha_s|^2+2|\alpha_i|^2)\alpha_h, \label{S9}
\end{align}
\end{widetext}
where $\omega = \omega_s-\omega_p/2$ is the signal detuning, $\Delta_b$ is defined in Eq.~\eqref{params1}, and $u_s$, $u_i$ and $u_h$ denote the drive strengths at corresponding frequencies, following the notations introduced in Ref.~\cite{Frattini2018}. Note that the pump amplitude $\alpha_p=u_p/\omega_a$ here does not include any corrections due to $\alpha_s$, $\alpha_i$ and $\alpha_h$ -- they have been reabsorbed into the above equations. 

To derive the boundaries in the stability diagram, we solve these equations in the small-signal approximation, in which the Stark shift due to $|\alpha_s|^2$ and $|\alpha_i|^2$ can be neglected. Since no input is sent at $\omega_p/2$, we can set the drive strength $u_h$ to zero. Then equation \eqref{S9} reduces to
\begin{align}
\big(\Delta_b-12g_4^*|\alpha_h|^2+i\kappa/2 \big)\alpha_h = 4g_3\alpha_p\alpha_h^*.
\end{align}
After multiplying it by the complex conjugate, we find a trivial solution $\alpha_h=0$ and possible high-amplitude solutions
\begin{align}
|\alpha_h|^2 = \frac{1}{12g_4^*}\left(\Delta_b\pm\sqrt{(4g_3|\alpha_p|)^2-\frac{\kappa^2}{4}}\right).\label{S_nontriv}
\end{align}

First, consider $\Delta_b>0$ half-plane. In this case only the solution with the minus sign in Eq.~\eqref{S_nontriv} is possible if the additional condition $(4g_3|\alpha_p|)^2-\frac{\kappa^2}{4}>\Delta_b^2$ is satisfied. This defines the separation line between regions (I) and (II) of the stability diagram. In region (I) only $\alpha_h=0$ solutions exists and is stable. High-amplitude solution exists in region (II) and is stable, while $\alpha_h=0$ is unstable.  

Next, consider $\Delta_b<0$ half-plane. The requirement for $|\alpha_h|^2$ being real leads to the condition $(4g_3|\alpha_p|)^2>~\frac{\kappa^2}{4}$. This condition defines the separation line between region (I) and (III) of the stability diagram. Below the separation line only the trivial solution is possible, while above this line both signs in Eq.~\eqref{S_nontriv} are possible. This yields five stationary points: a stable trivial $\alpha_h=0$ point, a pair of symmetric unstable points defined by the plus sign in Eq.~\eqref{S_nontriv} and a pair of symmetric stable points defined by the minus sign. The two unstable stationary points merge with the stable $\alpha_h=0$ point on the line which separates regions (II) and (III), defined by the equation $(4g_3|\alpha_p|)^2=\frac{\kappa^2}{4}+\Delta_b^2$. Above this line in region (II) the $\alpha_h=0$ point becomes unstable and the two symmetric high-amplitude states remain stable. 

The stability diagram plotted in coordinates $\Delta$ and $n_p$ (which directly translate into the experimental control knobs pump frequency $\omega_p$ and power $P_p$) is shown in Fig.~\ref{fig3}(a). Note that similar stability diagrams can be found in for example Refs.~\cite{Zorin2011,Wustmann2013}. 

Finally we note that by setting $\alpha_h=0$ (where appropriate, see Appendix~\ref{supp2}) the system of equations~\eqref{x1}-\eqref{x2} together with the input-output relation can be solved to yield the expression~\eqref{gain} for gain in the presence of a finite input signal.

\section{Further optimization of the SPA: an outlook  \label{supp5} }

Given the findings of this work, one can raise the question of how to improve the compression power of resonant parametric amplifiers beyond what has been achieved at a Kerr-free point in our particular device. Although sweet spots are in general useful, it is desirable to have a systematic approach to the problem of saturation of quantum-limited amplifiers. We would like to emphasize that such an approach was suggested in Ref.~\cite{Frattini2018}, where steady improvement of $P_{\rm 1dB}$ of an SPA was achieved by varying the design parameters in the direction corresponding to the increase of the coupling $\kappa$ and the decrease of the overall level of nonlinearity. 

Continuing to improve $P_{1\rm dB}$ in this fashion will require more pump power to be delivered to the device at the base of the dilution refrigerator. As such, it is important to compare state-of-the-art quantum-limited parametric amplifiers on the metric of power efficiency, here defined as the ratio of the signal output power at the 1 dB compression point to the required pump power $P_p$
\begin{align}
\eta_p = \frac{GP_{\rm 1 dB}}{P_p}.
\end{align}

As shown in Fig.~\ref{S4}, the amplifiers with off-resonant pumps, such as three-wave mixing SPAs and JPCs and flux-pumped JPAs, suffer from poor power efficiency, compared to the four-wave mixing amplifiers with near-resonant pumps and traveling-wave amplifiers. 

This is an important challenge on the way towards larger dynamic range. Indeed, as the required pump power increases, undesired effects such as  increased noise temperature or fridge heating, may result. An Oxford Triton dilution refrigerator will heat up by about $10\;\rm mK$ for $-20\rm\; dBm$ of applied power at the base stage. Typically, an attenuator is used to thermalize the pump to this stage. Assuming $20 \;\rm dB$ of attenuation, this puts the limit on allowable pump power at the plane of the device close to $-40\rm\; dBm$. This limit is fuzzy (shown as color gradient in Fig.~\ref{S4}), because it depends on the exact details of the attenuation and filtering chosen for the pump line. Nevertheless, it is clear that further increase of $P_{1\rm dB}$ at the expense of pump power is not a viable approach for optimization.

Instead, this issue needs to be addressed with better design of the pump coupling, which will allow to obtain the required $n_p$ with less pump power $P_p$ supplied to the plane of the device. Note that the large spectral separation of signal and pump in three-wave mixing amplifiers can be utilized to separately engineer the environmental properties seen by them. For example, the pump could be coupled through a separate resonant filter mode, which would also protect against the leakage of the signal through this port. At the same time, the signal port of the SPA can be optimized for bandwidth using the methods employed in Refs.~\cite{Mutus2014,Roy2015,Naaman2017}. These advances are left for future work.

\onecolumngrid

\section{Derivation of the SPA Hamiltonian  \label{supp6} }

The truncated Hamiltonian of the SPA, limited to only the fundamental mode $a$, further submitted to pumping, is given by
\begin{equation}
H_{{\rm SPA}}/\hbar=\omega_{a}a^{\dagger}a+g_{3}(a+a^{\dagger})^{3}+g_{4}(a+a^{\dagger})^{4}+...\label{eq:spa}
\end{equation}

The goal of this section is to obtain the dependence of the parameters $\omega_{a}$, $g_{3}$ and $g_{4}^*\equiv g_4-5g_3^2/\omega_a$ on the design parameters of the circuit. The expressions \eqref{eq:disper}, \eqref{g4} and \eqref{eq:g3} obtained here were used for fitting and making the design choices in this work and in Ref.~\cite{Frattini2018}.

\subsection*{Equations of motion in the distributed-element model of the SPA}

Let the length of each arm of the transmission line resonator be $d$, the capacitance to ground per unit length be $c$, and the inductance per unit length be $\ell$. We treat the array of $M$ SNAILs as a lumped circuit element located at $x=0$ with the phase distributed equally among all SNAILs in the array. We will denote the generalized flux field in the transmission line as $\phi(x)$. For simplicity, we also introduce the notation $\phi^{R/L}=\underset{x\to\pm0}{\lim}\phi(x)$. 

The Lagrangian can be written as
\begin{equation}
\mathcal L=\int_{-d}^{-0}\bigg[\frac{c}{2}(\partial_{t}\phi)^{2}-\frac{1}{2\ell}(\partial_{x}\phi)^{2}\bigg]dx+\int_{+0}^{d}\bigg[\frac{c}{2}(\partial_{t}\phi)^{2}-\frac{1}{2\ell}(\partial_{x}\phi)^{2}\bigg]dx-MU_{S}\bigg(\frac{\phi^{R}-\phi^{L}}{M\phi_{0}}\bigg),\label{eq:Lagrangian}
\end{equation}
where $\phi_{0}=\hbar/2e$ is the reduced flux quantum and $U_{S}(\varphi)$ is the potential energy of the SNAIL with a phase drop $\varphi$,
\begin{align}
U_S(\varphi) = -E_J\left[ \alpha\cos\varphi +3\cos\left(\frac{\varphi_{\rm ext}-\varphi}{3}\right) \right]
\end{align}
where $E_J$ is the Josephson energy of the large junctions, $\alpha$ is a ratio of the junction inductances in the two arms of the SNAIL, and $\varphi_{\rm ext}=2\pi\Phi/\Phi_0$ is the phase bias corresponding to the external magnetic flux $\Phi$, see \cite{Frattini2018}. 

The action for this Lagrangian is simply given by $S=\int_{t_{1}}^{t_{2}}\mathcal L(t)dt$, and can be split into three separate parts $S=S_{{\rm L}}+S_{{\rm R}}+S_{{\rm SNAIL}}$ corresponding to the separate contributions in the Lagrangian in Eq.~\ref{eq:Lagrangian}. The variation of action $\delta S$ contains three terms
\begin{align}
\delta S_{{\rm L}} & =-\int_{t_{1}}^{t_{2}}dt\bigg[\frac{1}{\ell}\partial_{x}\phi\bigg]\delta\phi\bigg|_{x=-d}^{x=-0}-\int_{t_{1}}^{t_{2}}dt\int_{-d}^{-0}\bigg[c\partial_{t}^{2}\phi-\frac{1}{\ell}\partial_{x}^{2}\phi\bigg]\delta\phi dx,\\
\delta S_{{\rm R}} & =-\int_{t_{1}}^{t_{2}}dt\bigg[\frac{1}{\ell}\partial_{x}\phi\bigg]\delta\phi\bigg|_{x=+0}^{x=d}-\int_{t_{1}}^{t_{2}}dt\int_{+0}^{d}\bigg[c\partial_{t}^{2}\phi-\frac{1}{\ell}\partial_{x}^{2}\phi\bigg]\delta\phi dx,\\
\delta S_{{\rm SNAIL}} & =-M\int_{t_{1}}^{t_{2}}dt\frac{\partial U_{S}(\varphi)}{\partial\varphi}\frac{\delta\phi^{R}-\delta\phi^{L}}{M\phi_{0}}\bigg|_{\varphi=\frac{\phi^{R}-\phi^{L}}{M\phi_{0}}}.
\end{align}

According to the variational principle, by requiring $\delta S=0$ we obtain the equations of motion for the flux $\phi(x,t)$ and the boundary conditions. 
\begin{itemize}[leftmargin=*]
\item[$\circ$] In the bulk of the transmission line resonator, for $x\in(-d,0)$ and $x\in(0,d)$, the flux field $\phi(x)$ obeys the wave equation
\begin{equation}
\partial_{t}^{2}\phi-v^{2}\partial_{x}^{2}\phi=0.\label{eq:Klein-Gordon}
\end{equation}
where $v=1/\sqrt{\ell c}$ is the phase velocity.
\item[$\circ$] At the boundaries $x=-d$ and $x=d$ the flux field has zero current boundary condition
\begin{equation}
\partial_{x}\phi=0.\label{eq:boundary-1}
\end{equation}
\item[$\circ$] In the center of the resonator, where it is interrupted by the SNAIL array, the left and right continuity conditions should hold
\begin{equation}
\frac{1}{\phi_0}\frac{\partial U_{S}(\varphi)}{\partial\varphi}\bigg|_{\varphi=\frac{\phi^{R}-\phi^{L}}{M\phi_{0}}}=\frac{1}{\ell}\partial_{x}\phi^{L/R},\label{continuity_1}
\end{equation}
which also implies the current conservation $\partial_{x}\phi^{L}=\partial_{x}\phi^{R}$ across the array. Strictly speaking, the bare SNAIL potential $U_{S}(\varphi)$ is $6\pi$-periodic in $\varphi$. However, we are interested in a time scale on which the phase slips can be neglected, and, thus, the fact that $\varphi$ is compact does not matter. In this approximation, we can assume that the phase $\varphi$ is localized near the bottom of one of the many equivalent potential minima and expand the potential around that point (denoted as $\varphi_{\rm min}$) 
\begin{equation}
U_{S}(\varphi)=E_{J}\bigg(\frac{c_{2}}{2!}(\varphi-\varphi_{{\rm min}})^{2}+\frac{c_{3}}{3!}(\varphi-\varphi_{{\rm min}})^{3}+\frac{c_{4}}{4!}(\varphi-\varphi_{{\rm min}})^{4}+...\bigg),
\end{equation}
where all dimensionless expansion coefficients $c_{k}\equiv c_{k}(\Phi,\alpha)$ depend only on the external magnetic flux $\Phi$ and junction inductance ratio $\alpha$. Using this expansion truncated to the fourth order, we can rewrite the continuity condition at $x=\pm0$ as
\begin{equation}
\frac{E_{J}}{\phi_0}\bigg[c_{2}\left(\frac{\phi^{R}-\phi^{L}}{M\phi_{0}}-\varphi_{{\rm min}}\right)+\frac{c_{3}}{2!}\left(\frac{\phi^{R}-\phi^{L}}{M\phi_{0}}-\varphi_{{\rm min}}\right)^{2}+\frac{c_{4}}{3!}\left(\frac{\phi^{R}-\phi^{L}}{M\phi_{0}}-\varphi_{{\rm min}}\right)^{3}\bigg]=\frac{1}{\ell}\partial_{x}\phi^{L/R}.\label{continuity}
\end{equation}
This condition will allow us to match the solutions of the wave equation \eqref{eq:Klein-Gordon} in the left and right arms of the microstrip transmission line. 
\end{itemize}
Instead of working with the continuous flux field $\phi(x)$, it is convenient to decompose it using the eigenmode excitations of the circuit. This is done by using as a basis the general solution to Eq.~\eqref{eq:Klein-Gordon} in the left and right arms of the transmission line resonator
\begin{align}
\phi(x,t) & =(a_{0}^{L/R}+b_{0}^{L/R}x)+\sum_{n}\bigg[a_{n}^{L/R}\cos\bigg(\frac{\omega_{n}x}{v}\bigg)+b_{n}^{L/R}\sin\bigg(\frac{\omega_{n}x}{v}\bigg)\bigg],\label{eq:field}\\
\phi^{L/R} & =a_{0}^{L/R}+\sum_{n}a_{n}^{L/R},\label{eq:left right}
\end{align}
where the coefficients $a_{n}^{L/R}(t)$ and $b_{n}^{L/R}(t)$ should satisfy $[\partial_{t}^{2}+\omega_{n}^{2}](\cdot)=0$. We need to rewrite the boundary condition in Eq.~\eqref{eq:boundary-1} and the continuity condition in Eq.~\eqref{continuity} using the eigenmode decomposition. Starting with the boundary condition, by differentiating Eq.~\eqref{eq:field} we obtain
\begin{equation}
b_{0}^{L/R}+\sum_{n}\frac{\omega_{n}}{v}\bigg[\pm a_{n}^{L/R}\sin\bigg(\frac{\omega_{n}d}{v}\bigg)+b_{n}^{L/R}\cos\bigg(\frac{\omega_{n}d}{v}\bigg)\bigg]=0,
\end{equation}
and since this has to be satisfied at all times, applying harmonic balance leads to
\begin{equation}
\pm a_{n}^{L/R}\sin\bigg(\frac{\omega_{n}d}{v}\bigg)+b_{n}^{L/R}\cos\bigg(\frac{\omega_{n}d}{v}\bigg)=0,\label{boundary}
\end{equation}
and also $b_{0}^{L/R}=0$, meaning that there is no uniform static flux gradient in the resonator. Now we can similarly treat the continuity condition in Eq.~\eqref{continuity} using Eq.~\eqref{eq:left right} and its derivative
\begin{equation}
\partial_{x}\phi^{L/R}=b_{0}^{L/R}+\sum_{n}\frac{\omega_{n}}{v}b_{n}^{L/R}.\label{eq:left right-1}
\end{equation}

The continuity condition contains two equations for the left and right side of the array. Equivalently, we can use one of them and the current conservation condition $\partial_{x}\phi^{L}=\partial_{x}\phi^{R}$ instead. With the help of Eq.~\eqref{eq:left right-1} and applying harmonic balance, this reduces to
\begin{equation}
b_{n}^{L}=b_{n}^{R}.\label{boundary3}
\end{equation}

As the second continuity equation we choose Eq.~\eqref{continuity} on the right side of the array, which with the help of Eqs.~\eqref{eq:field} and \eqref{eq:left right-1} can be transformed to
\begin{equation}
\frac{E_{J}}{\phi_0}\left[c_{2}\left(\sum_{n=0}^{\infty}\frac{a_{n}^{R}-a_{n}^{L}}{M\phi_{0}}-\varphi_{{\rm min}}\right)+\frac{c_{3}}{2!}\left(\sum_{n=0}^{\infty}\frac{a_{n}^{R}-a_{n}^{L}}{M\phi_{0}}-\varphi_{{\rm min}}\right)^{2}+\frac{c_{4}}{3!}\left(\sum_{n=0}^{\infty}\frac{a_{n}^{R}-a_{n}^{L}}{M\phi_{0}}-\varphi_{{\rm min}}\right)^{3}\right]-\frac{1}{\ell}\sum_{n}\frac{\omega_{n}}{v}b_{n}^{R}=0.\label{nonline}
\end{equation}

Note that this equation is nonlinear, which leads to the coupling of various eigenmodes by the SNAIL array mixer. In the limit of weak excitations, we can linearize the system and decompose it into a collection of harmonic oscillators with eigenfrequencies $\omega_{n}$. The lowest such eigenmode of the device is used to implement the SPA.

\subsection*{Eigenmode decomposition}

The linear part of Eq.~\eqref{nonline} is simply
\[
\frac{1}{L_{s}}\left(\sum_{n=0}^{\infty}\frac{a_{n}^{R}-a_{n}^{L}}{M}-\phi_{{\rm min}}\right)-\frac{1}{\ell}\sum_{n}\frac{\omega_{n}}{v}b_{n}^{R}=0,
\]
where we have introduced the inductance of the large junction of a SNAIL $L_{J}=\varphi_{0}/E_{J}$, the flux-dependent SNAIL inductance $L_{s}(\Phi)=L_{J}/c_{2}(\Phi)$ and the flux offset $\phi_{\rm min}=\varphi_{\rm min}\,\phi_0$. After applying harmonic balance to this equation we obtain
\begin{align}
\frac{a_{0}^{R}-a_{0}^{L}}{M}-\phi_{{\rm min}} & =0,\label{flux offset}\\
\frac{1}{L_{s}}\frac{a_{n}^{R}-a_{n}^{L}}{M}-\frac{1}{\ell}\frac{\omega_{n}}{v}b_{n}^{R} & =0.\label{eq:boundary2}
\end{align}

Now the eigenmodes of the circuit can be found using the equations \eqref{boundary}, \eqref{boundary3} and \eqref{eq:boundary2}. It is convenient to combine them in a matrix form
\begin{equation}
\left[\begin{array}{cccc}
\tan\big(\frac{\omega_{n}d}{v}\big) & 1 & 0 & 0\\
0 & 0 & -\tan\big(\frac{\omega_{n}d}{v}\big) & 1\\
0 & 1 & 0 & -1\\
-1 & 0 & 1 & -\frac{L_{s}M\omega_{n}}{Z_{c}}
\end{array}\right]\left(\begin{array}{c}
a_{n}^{L}\\
b_{n}^{L}\\
a_{n}^{R}\\
b_{n}^{R}
\end{array}\right)=0,\label{eq:matrix equation}
\end{equation}
where we have introduced the characteristic impedance of the transmission line resonator $Z_{c}=\sqrt{\ell/c}$. By requiring that the determinant of Eq.~\eqref{eq:matrix equation} be zero we obtain the equation for eigenfrequencies, while the eigenvectors define the coefficients needed to calculate the flux profile of the eigenmodes in Eq.~\eqref{eq:field}. The equation for eigenfrequencies is
\begin{equation}
\frac{2Z_{c}}{ML_{s}(\Phi)}=\omega_n\tan\bigg(\frac{\pi}{2}\frac{\omega_{n}}{\omega_{0}}\bigg),\label{eq:disper}
\end{equation}
where we have explicitly indicated the flux dependence of the SNAIL inductance and defined the parameter $\omega_{0}=\frac{\pi}{2}\frac{v}{d}$, whose physical meaning is the frequency of the fundamental harmonic of the resonator in the absence of the array. 

We will use the value of the flux field at the right boundary $\phi_{n}(t)=\phi_{n}(d,t)$ as the canonical coordinate for each mode. It is a natural choice, since this variable is the one to which we couple the input transmission line via a weak coupling capacitor. In terms of $\phi_{n}$ we can write the components of the eigenvector of Eq.\eqref{eq:matrix equation} as
\begin{equation}
a_{n}^{R}=\phi_{n}\cos\bigg(\frac{\pi}{2}\frac{\omega_{n}}{\omega_{0}}\bigg),\qquad a_{n}^{L}=\phi_{n}\bigg[\cos\bigg(\frac{\pi}{2}\frac{\omega_{n}}{\omega_{0}}\bigg)-\frac{ML_{s}\omega_{n}}{Z_{c}}\sin\bigg(\frac{\pi}{2}\frac{\omega_{n}}{\omega_{0}}\bigg)\bigg],\qquad b_{n}^{L/R}=\phi_{n}\sin\bigg(\frac{\pi}{2}\frac{\omega_{n}}{\omega_{0}}\bigg),
\end{equation}
and the flux distribution in each eigenmode as
\begin{equation}
\phi_{n}(x,t)=\phi_{n}\bigg[{\rm sgn\,}x\cos\bigg(\frac{\omega_{n}x}{v}\bigg)\cos\bigg(\frac{\pi}{2}\frac{\omega_{n}}{\omega_{0}}\bigg)+\sin\bigg(\frac{\pi}{2}\frac{\omega_{n}}{\omega_{0}}\bigg)\sin\bigg(\frac{\omega_{n}x}{v}\bigg)\bigg].\label{eq:flux profile}
\end{equation}

Note also that the constant flux offset between the two arms of the resonator is given by Eq.~\eqref{flux offset}. After substituting the flux distribution given by Eq.~\eqref{eq:field} into the Lagrangian and calculating the $x$-integral, we can diagonalize it and reduce to the collection of independent harmonic oscillators
\begin{equation}
\mathcal L=\sum_{n}\bigg(\frac{C_{n}\dot{\phi}_{n}^{2}}{2}-\frac{\phi_{n}^{2}}{2L_{n}}\bigg),
\end{equation}
with the mode capacitance and inductance given by
\begin{align}
C_{n} & =\frac{1}{2\omega_{n}Z_{c}}\bigg[\pi\frac{\omega_{n}}{\omega_{0}}+\sin\bigg(\pi\frac{\omega_{n}}{\omega_{0}}\bigg)\bigg],\label{eq:cap}\\
L_{n}^{-1} & =\frac{\omega_{n}}{2Z_{c}}\bigg[\pi\frac{\omega_{n}}{\omega_{0}}+\sin\bigg(\pi\frac{\omega_{n}}{\omega_{0}}\bigg)\bigg],\label{eq:ind}
\end{align}
and the mode resonant frequency $\omega_{n}=1/\sqrt{L_{n}C_{n}}$ and impedance $Z_n=\sqrt{L_n/C_n}$. Note that in general both $C_{n}$ and $L_{n}$ can depend on $\Phi$. 

\subsection*{Nonlinearities of the SPA}

Having considered the linear properties of the SPA resonator, we will now focus on deriving the dependence of nonlinearities $g_3$ and $g_4^*$ on the design parameters of the circuit. Such a perturbative approach, with the first step consisting of solving the linearized problem, is valid if $Z_n\ll R_Q$ \cite{Manucharyan2012}, where $Z_n$ is the mode impedance and $R_{Q}=\hbar/(2e)^{2}$ is the reduced resistance quantum.

Because of the nonlinearity of the array, exciting a particular nonlinear mode $n$ leads to the oscillations not only at the frequency $\omega_{n}$, but also at higher harmonics of this frequency $2\omega_{n}$, $3\omega_{n}$ etc.  The amplitudes of the oscillations at the multiples of the mode frequency $\omega_n$ are small if the mode is not significantly excited, and if the system is weakly nonlinear.  This  allows to solve the problem in a perturbative manner, similar to the calculation in \cite{Wallquist2006}, using the eigenmodes found in the previous section as the first approximation. 

In this section, we focus on the lowest mode $n=1$ of the circuit. This single-mode approximation is justified in the SPA resonator for two reasons. First, the SPA mode is in the nearly ``lumped'' regime, where almost all inductance comes from the array and all capacitance comes from the microstrip pads. We have experimentally verified using two-tone spectroscopy that the next mode of the SPA is at $25\rm \; GHz$ (data not shown). Since the detuning of the higher modes is much larger than the lowest mode frequency, we expect the effect of these modes on the nonlinearities of the lowest mode to be small. Second, all the driving signals in our experiments are significantly below these higher modes, and,  therefore, do not excite them.

To be consistent with Eq.~\eqref{eq:spa}, we will denote the frequency of the mode as $\omega_{a}$ and canonical coordinate as $\phi_{a}$. The flux profile of the excited mode should still obey the equation \eqref{eq:Klein-Gordon} in the left and right arms of the transmission line resonator, and, thus, we can write it as
\begin{equation}
\phi_{a}(x,t)=u_{0}^{L/R}+\sum_{k}\bigg[u_{k}^{L/R}\cos\bigg(\frac{k\omega_{a}x}{v}\bigg)+v_{k}^{L/R}\sin\bigg(\frac{k\omega_{a}x}{v}\bigg)\bigg],
\end{equation}
instead of Eq.~\eqref{eq:flux profile}. The new coefficients $u_{k}^{L/R}(t)$ and $v_{k}^{L/R}(t)$ should obey the equation $[\partial_{t}^{2}+(n\omega_{a})^{2}](\cdot)=0$. In the linear approximation, we have already found the coefficients of this expansion in the previous section. The DC part obeys $u_{0}^{R}-u_{0}^{L}=M\phi_{{\rm min}}$, and the only nonzero coefficients are $u_{1}^{R/L}=\pm\phi_{a}\cos\big(\frac{\pi}{2}\frac{\omega_{a}}{\omega_{0}}\big)$ and $v_{1}^{R/L}=\phi_{a}\sin\big(\frac{\pi}{2}\frac{\omega_{a}}{\omega_{0}}\big)$;
the rest of the harmonics with $k>1$ are not excited. Using the perturbative approach, we will find that the amplitudes of these higher harmonics are suppressed, for example we will find that $u_{2}^{R/L}\propto\phi_{a}^{2}$. But first, by using the boundary conditions and harmonic balance we can establish a useful relation between $u_{k}^{L/R}$ and $v_{k}^{L/R}$, which will allow us to reduce the number of independent coefficients. This way we find that $u_{k}^{L}=-u_{k}^{R}$ and $v_{k}^{R}=u_{k}^{R}\tan\big(\frac{\pi}{2}\frac{k\omega_{a}}{\omega_{0}}\big)$, and thus for each $k$ we only need to use $u_{k}^{R}$. Let us now return to the continuity condition in Eq.~\eqref{nonline}, which we have previously linearized, but this time treat it as nonlinear. We will set the amplitudes of all higher modes with $n>1$ to zero, but include the higher harmonics of the first mode, truncated at the second order. Then Eq.~\eqref{nonline} reduces to
\begin{equation}
\sum_{m=1}^{3}\frac{c_{m+1}}{\phi_{0}^{m-1}m!}\left(\sum_{k=1}^{2}\frac{2u_{k}^{R}}{M}+\frac{u_{0}^{R}-u_{0}^{L}}{M}-\phi_{{\rm min}}\right)^{m}-\frac{\omega L_{J}}{Z_{c}}\sum_{k=1}^{2}ku_{k}^{R}\tan\bigg(\frac{\pi}{2}\frac{k\omega_{a}}{\omega_{0}}\bigg)=0.\label{eq:huge formula}
\end{equation}

By including the time dependence $u_{1}^{R}(t)=u_{1}^{R}\sin\omega_{a}t$ and $u_{2}^{R}(t)=u_{2}^{R}\cos2\omega_{a}t$   and keeping the relevant contributions (in the spirit of perturbation theory) we solve Eq.~\eqref{eq:huge formula} by harmonic balance. 
\begin{itemize}[leftmargin=*]
\item[$\circ$] DC harmonic
\begin{equation}
\frac{u_{0}^{R}-u_{0}^{L}}{M}-\phi_{{\rm min}}=-\frac{c_{3}}{c_{2}}\frac{1}{M^{2}}\frac{(u_{1}^{R})^{2}}{\phi_{0}}. \label{u0}
\end{equation}
\item[$\circ$] Second harmonic $2\omega_{a}$
\begin{equation}
u_{2}^{R}=\frac{c_{3}}{c_{2}-\frac{\omega_{a}L_{J}M}{Z_{c}}\tan\big(\pi\frac{\omega_{a}}{\omega_{0}}\big)}\frac{1}{2M}\frac{(u_{1}^{R})^{2}}{\phi_{0}}. \label{u2}
\end{equation}
\item[$\circ$] First harmonic $\omega_{a}$
\begin{equation}
\bigg[\frac{2c_{2}}{M}-\frac{\omega_{a}L_{J}}{Z_{c}}\tan\bigg(\frac{\pi}{2}\frac{\omega_{a}}{\omega_{0}}\bigg)\bigg]+\bigg[c_{4}-\frac{2c_{3}^{2}}{c_{2}}-\frac{c_{3}^{2}}{c_{2}-\frac{\omega_{a}L_{J}M}{Z_{c}}\tan\big(\pi\frac{\omega_{a}}{\omega_{0}}\big)}\bigg]\frac{1}{M^{3}}\frac{(u_{1}^{R})^{2}}{\phi_{0}^{2}}=0.\label{eq:dispersion}
\end{equation}
Note that to write the relation \eqref{eq:dispersion} we had to use the expressions~\eqref{u0} and \eqref{u2} to eliminate $u_0^R$, $u_0^L$ and $u_2^R$.
\end{itemize}
The equation \eqref{eq:dispersion} is essentially equivalent to the dispersion relation Eq.~\eqref{eq:disper}, corrected by an additional term proportional to the intensity of the excitation -- a result of the Kerr effect. By subtracting Eq.~\eqref{eq:disper} from Eq.~\eqref{eq:dispersion} and replacing the finite difference with the derivative we obtain
\begin{equation}
\Delta\omega=\frac{Z_{c}}{2L_{J}}\frac{\cos^{2}\big(\frac{\pi}{2}\frac{\omega_{a}}{\omega_{0}}\big)\sin\big(\pi\frac{\omega_{a}}{\omega_{0}}\big)}{\frac{\pi}{\omega_{0}}\frac{Z_{c}}{ML_{s}}+\sin^{2}\big(\frac{\pi}{2}\frac{\omega_{a}}{\omega_{0}}\big)}\frac{1}{M^{3}}\bigg[c_{4}-\frac{2c_{3}^{2}}{c_{2}}-\frac{c_{3}^{2}}{c_{2}-\frac{\omega_{a}L_{J}M}{Z_{c}}\tan\big(\pi\frac{\omega_{a}}{\omega_{0}}\big)}\bigg]\frac{\phi_{a}^{2}}{\phi_{0}^{2}},\label{eq:shift}
\end{equation}
where we have used the relation $u_{1}^{R}=\phi_{a}\cos\big(\frac{\pi}{2}\frac{\omega_{a}}{\omega_{0}}\big)$ to express the resulting frequency shift $\Delta\omega$ in terms of the canonical coordinate $\phi_a$. 

We can also compute the linear shift of the resonant frequency due to the finite average mode population $\overline{n}$ using the model described by Eq.~\eqref{eq:spa}, which results in $\Delta\omega=12g_{4}^{*}\overline{n}$ with $g_{4}^{*}=g_{4}-5g_{3}^{2}/\omega_{a}$. To directly compare these two results for $\Delta\omega$, we need to express $\phi_{a}^{2}$ in terms of the average photon number as well, which can be done using the conventional method of canonical quantization. This leads to $\phi_{a}=\phi_{{\rm ZPF}}(a+a^{\dagger})$, where $\phi_{{\rm ZPF}}=\sqrt{\hbar Z_{1}/2}$ and $Z_{1}=\sqrt{L_{1}/C_{1}}$ is the impedance of the mode calculated with the help of Eqs.~\eqref{eq:cap} and \eqref{eq:ind}. Comparison of the two expressions for the shift $\Delta\omega$ leads to the following
\begin{equation}
g_{4}^{*}=\frac{1}{12}\frac{\omega_{a}\sin^{2}\big(\pi\frac{\omega_{a}}{\omega_{0}}\big)}{c_{2}M^{2}\tan\big(\frac{\pi}{2}\frac{\omega_{a}}{\omega_{0}}\big)\big[\pi\frac{\omega_{a}}{\omega_{0}}+\sin\big(\pi\frac{\omega_{a}}{\omega_{0}}\big)\big]^{2}}\frac{Z_{c}}{R_{Q}}\bigg[c_{4}-\frac{c_{3}^{2}}{c_{2}}\frac{3+5\big(\frac{\omega_{a}ML_{s}}{2Z_{c}}\big)^{2}}{1+3\big(\frac{\omega_{a}ML_{s}}{2Z_{c}}\big)^{2}}\bigg],\label{g4}
\end{equation}

The derivation of $g_4^*$ presented here relies crucially on the careful treatment of higher harmonics of the excited mode. Such treatment is necessary for any higher-order nonlinearity of the system. However, to find the lowest-order nonlinearity $g_{3}$, we can employ the traditional black-box quantization method \cite{Nigg2012}. In this method, the circuit is first linearized, and the nonlinearity is then applied to the phase drop across the nonlinear elements of the original circuit. Therefore, the cubic contribution to the Hamiltonian in Eq.~\eqref{eq:spa} coming from the array can be written simply as
\begin{equation}
ME_{J}\frac{c_{3}}{3!}\left(\frac{\phi^{R}-\phi^{L}}{M\phi_{0}}-\varphi_{{\rm min}}\right)^{3}\to\frac{\phi_{a}^{3}}{\phi_{0}}\frac{8}{M^{2}L_{J}}\frac{c_{3}}{3!}\cos^{3}\bigg(\frac{\pi}{2}\frac{\omega_{a}}{\omega_{0}}\bigg).
\end{equation}

Plugging in the expression $\phi_{a}=\phi_{{\rm ZPF}}(a+a^{\dagger})$, we recover the cubic term $g_{3}(a+a^{\dagger})^{3}$ with $g_{3}$ given by
\begin{equation}
g_{3}=\frac{4Z_{c}c_{3}}{3M^{2}L_{J}}\bigg(\frac{\cos^{2}\big(\frac{\pi}{2}\frac{\omega_{a}}{\omega_{0}}\big)}{\pi\frac{\omega_{a}}{\omega_{0}}+\sin\big(\pi\frac{\omega_{a}}{\omega_{0}}\big)}\bigg)^{3/2}\sqrt{\frac{Z_{c}}{R_{Q}}}.\label{eq:g3}
\end{equation}

\subsection*{Fitting}

To conclude this section, we have calculated the coefficients $\omega_a({\cal S},\Phi)$, $g_3({\cal S},\Phi)$ and $g_4({\cal S},\Phi)$ of the SPA Hamiltonian as functions of the design parameters ${\cal S}=\{M, \omega_0,Z_c,L_J,\alpha\}$ of the device, as well as the external magnetic flux $\Phi$. In the experiment, we fit the resonance frequency $\omega_a(\Phi)$ in order to refine the knowledge of the parameters in $\cal S$. Those are obtained in the following way: $Z_{c}=45.8{\rm\:\Omega}$ is defined by the microstrip geometry, $L_{J}=38\,{\rm pH}$ is calculated using the room-temperature measurements of junction's resistance and the Ambegaokar-Baratoff formula, $M=20$ is the number of SNAILs in the array, and $\omega_{0}=2\pi\times16.0\,{\rm GHz}$ and $\alpha=0.065$ are fitted and are found to be close to their design values. The measurement results of $g_{4}^{*}$ and $g_{3}$ are compared to the predicted values using the refined parameters $\cal S$ in Fig.~\ref{fig1}(b). 

\twocolumngrid
\bibliographystyle{apsrev_longbib}
\bibliography{library}

\begin{thebibliography}{58}
\expandafter\ifx\csname natexlab\endcsname\relax\def\natexlab#1{#1}\fi
\expandafter\ifx\csname bibnamefont\endcsname\relax
  \def\bibnamefont#1{#1}\fi
\expandafter\ifx\csname bibfnamefont\endcsname\relax
  \def\bibfnamefont#1{#1}\fi
\expandafter\ifx\csname citenamefont\endcsname\relax
  \def\citenamefont#1{#1}\fi
\expandafter\ifx\csname url\endcsname\relax
  \def\url#1{\texttt{#1}}\fi
\expandafter\ifx\csname urlprefix\endcsname\relax\def\urlprefix{URL }\fi
\providecommand{\bibinfo}[2]{#2}
\providecommand{\eprint}[2][]{\url{#2}}

\bibitem[{\citenamefont{Clerk et~al.}(2010)\citenamefont{Clerk, Devoret,
  Girvin, Marquardt, and Schoelkopf}}]{Clerk2008}
\bibinfo{author}{\bibfnamefont{A.~A.} \bibnamefont{Clerk}},
  \bibinfo{author}{\bibfnamefont{M.~H.} \bibnamefont{Devoret}},
  \bibinfo{author}{\bibfnamefont{S.~M.} \bibnamefont{Girvin}},
  \bibinfo{author}{\bibfnamefont{F.}~\bibnamefont{Marquardt}},
  \bibnamefont{and} \bibinfo{author}{\bibfnamefont{R.~J.}
  \bibnamefont{Schoelkopf}}, \emph{\bibinfo{title}{{Introduction to quantum
  noise, measurement, and amplification}}}, \bibinfo{journal}{Reviews of Modern
  Physics} \textbf{\bibinfo{volume}{82}}, \bibinfo{pages}{1155}
  (\bibinfo{year}{2010}).

\bibitem[{\citenamefont{Roy and Devoret}(2016)}]{Devoret2016}
\bibinfo{author}{\bibfnamefont{A.}~\bibnamefont{Roy}} \bibnamefont{and}
  \bibinfo{author}{\bibfnamefont{M.}~\bibnamefont{Devoret}},
  \emph{\bibinfo{title}{{Introduction to parametric amplification of quantum
  signals with Josephson circuits}}}, \bibinfo{journal}{Comptes Rendus
  Physique} \textbf{\bibinfo{volume}{17}}, \bibinfo{pages}{740}
  (\bibinfo{year}{2016}).

\bibitem[{\citenamefont{Axline et~al.}(2018)\citenamefont{Axline, Burkhart,
  Pfaff, Zhang, Chou, Campagne-Ibarcq, Reinhold, Frunzio, Girvin, Jiang
  et~al.}}]{Axline2018}
\bibinfo{author}{\bibfnamefont{C.~J.} \bibnamefont{Axline}},
  \bibinfo{author}{\bibfnamefont{L.~D.} \bibnamefont{Burkhart}},
  \bibinfo{author}{\bibfnamefont{W.}~\bibnamefont{Pfaff}},
  \bibinfo{author}{\bibfnamefont{M.}~\bibnamefont{Zhang}},
  \bibinfo{author}{\bibfnamefont{K.}~\bibnamefont{Chou}},
  \bibinfo{author}{\bibfnamefont{P.}~\bibnamefont{Campagne-Ibarcq}},
  \bibinfo{author}{\bibfnamefont{P.}~\bibnamefont{Reinhold}},
  \bibinfo{author}{\bibfnamefont{L.}~\bibnamefont{Frunzio}},
  \bibinfo{author}{\bibfnamefont{S.~M.} \bibnamefont{Girvin}},
  \bibinfo{author}{\bibfnamefont{L.}~\bibnamefont{Jiang}},
  \bibnamefont{et~al.}, \emph{\bibinfo{title}{{On-demand quantum state transfer
  and entanglement between remote microwave cavity memories}}},
  \bibinfo{journal}{Nature Physics} \textbf{\bibinfo{volume}{14}},
  \bibinfo{pages}{705} (\bibinfo{year}{2018}).

\bibitem[{\citenamefont{Campagne-Ibarcq
  et~al.}(2018)\citenamefont{Campagne-Ibarcq, Zalys-Geller, Narla, Shankar,
  Reinhold, Burkhart, Axline, Pfaff, Frunzio, Schoelkopf
  et~al.}}]{Campagne-Ibarcq2018}
\bibinfo{author}{\bibfnamefont{P.}~\bibnamefont{Campagne-Ibarcq}},
  \bibinfo{author}{\bibfnamefont{E.}~\bibnamefont{Zalys-Geller}},
  \bibinfo{author}{\bibfnamefont{A.}~\bibnamefont{Narla}},
  \bibinfo{author}{\bibfnamefont{S.}~\bibnamefont{Shankar}},
  \bibinfo{author}{\bibfnamefont{P.}~\bibnamefont{Reinhold}},
  \bibinfo{author}{\bibfnamefont{L.}~\bibnamefont{Burkhart}},
  \bibinfo{author}{\bibfnamefont{C.}~\bibnamefont{Axline}},
  \bibinfo{author}{\bibfnamefont{W.}~\bibnamefont{Pfaff}},
  \bibinfo{author}{\bibfnamefont{L.}~\bibnamefont{Frunzio}},
  \bibinfo{author}{\bibfnamefont{R.~J.} \bibnamefont{Schoelkopf}},
  \bibnamefont{et~al.}, \emph{\bibinfo{title}{{Deterministic Remote
  Entanglement of Superconducting Circuits through Microwave Two-Photon
  Transitions}}}, \bibinfo{journal}{Physical Review Letters}
  \textbf{\bibinfo{volume}{120}}, \bibinfo{pages}{200501}
  (\bibinfo{year}{2018}).

\bibitem[{\citenamefont{Kurpiers et~al.}(2018)\citenamefont{Kurpiers, Magnard,
  Walter, Royer, Pechal, Heinsoo, Salath{\'{e}}, Akin, Storz, Besse
  et~al.}}]{Kurpiers2018}
\bibinfo{author}{\bibfnamefont{P.}~\bibnamefont{Kurpiers}},
  \bibinfo{author}{\bibfnamefont{P.}~\bibnamefont{Magnard}},
  \bibinfo{author}{\bibfnamefont{T.}~\bibnamefont{Walter}},
  \bibinfo{author}{\bibfnamefont{B.}~\bibnamefont{Royer}},
  \bibinfo{author}{\bibfnamefont{M.}~\bibnamefont{Pechal}},
  \bibinfo{author}{\bibfnamefont{J.}~\bibnamefont{Heinsoo}},
  \bibinfo{author}{\bibfnamefont{Y.}~\bibnamefont{Salath{\'{e}}}},
  \bibinfo{author}{\bibfnamefont{A.}~\bibnamefont{Akin}},
  \bibinfo{author}{\bibfnamefont{S.}~\bibnamefont{Storz}},
  \bibinfo{author}{\bibfnamefont{J.-C.} \bibnamefont{Besse}},
  \bibnamefont{et~al.}, \emph{\bibinfo{title}{{Deterministic quantum state
  transfer and remote entanglement using microwave photons}}},
  \bibinfo{journal}{Nature} \textbf{\bibinfo{volume}{558}},
  \bibinfo{pages}{264} (\bibinfo{year}{2018}).

\bibitem[{\citenamefont{Sliwa et~al.}(2015)\citenamefont{Sliwa, Hatridge,
  Narla, Shankar, Frunzio, Schoelkopf, and Devoret}}]{Sliwa2015}
\bibinfo{author}{\bibfnamefont{K.~M.} \bibnamefont{Sliwa}},
  \bibinfo{author}{\bibfnamefont{M.}~\bibnamefont{Hatridge}},
  \bibinfo{author}{\bibfnamefont{A.}~\bibnamefont{Narla}},
  \bibinfo{author}{\bibfnamefont{S.}~\bibnamefont{Shankar}},
  \bibinfo{author}{\bibfnamefont{L.}~\bibnamefont{Frunzio}},
  \bibinfo{author}{\bibfnamefont{R.~J.} \bibnamefont{Schoelkopf}},
  \bibnamefont{and} \bibinfo{author}{\bibfnamefont{M.~H.}
  \bibnamefont{Devoret}}, \emph{\bibinfo{title}{{Reconfigurable Josephson
  Circulator/Directional Amplifier}}}, \bibinfo{journal}{Physical Review X}
  \textbf{\bibinfo{volume}{5}}, \bibinfo{pages}{041020} (\bibinfo{year}{2015}).

\bibitem[{\citenamefont{Lecocq et~al.}(2017)\citenamefont{Lecocq, Ranzani,
  Peterson, Cicak, Simmonds, Teufel, and Aumentado}}]{Lecocq2017}
\bibinfo{author}{\bibfnamefont{F.}~\bibnamefont{Lecocq}},
  \bibinfo{author}{\bibfnamefont{L.}~\bibnamefont{Ranzani}},
  \bibinfo{author}{\bibfnamefont{G.~A.} \bibnamefont{Peterson}},
  \bibinfo{author}{\bibfnamefont{K.}~\bibnamefont{Cicak}},
  \bibinfo{author}{\bibfnamefont{R.~W.} \bibnamefont{Simmonds}},
  \bibinfo{author}{\bibfnamefont{J.~D.} \bibnamefont{Teufel}},
  \bibnamefont{and}
  \bibinfo{author}{\bibfnamefont{J.}~\bibnamefont{Aumentado}},
  \emph{\bibinfo{title}{{Nonreciprocal Microwave Signal Processing with a
  Field-Programmable Josephson Amplifier}}}, \bibinfo{journal}{Physical Review
  Applied} \textbf{\bibinfo{volume}{7}}, \bibinfo{pages}{024028}
  (\bibinfo{year}{2017}).

\bibitem[{\citenamefont{Fang et~al.}(2012)\citenamefont{Fang, Yu, and
  Fan}}]{Fang2012}
\bibinfo{author}{\bibfnamefont{K.}~\bibnamefont{Fang}},
  \bibinfo{author}{\bibfnamefont{Z.}~\bibnamefont{Yu}}, \bibnamefont{and}
  \bibinfo{author}{\bibfnamefont{S.}~\bibnamefont{Fan}},
  \emph{\bibinfo{title}{{Realizing effective magnetic field for photons by
  controlling the phase of dynamic modulation}}}, \bibinfo{journal}{Nature
  Photonics} \textbf{\bibinfo{volume}{6}}, \bibinfo{pages}{782}
  (\bibinfo{year}{2012}).

\bibitem[{\citenamefont{Peano et~al.}(2016)\citenamefont{Peano, Houde,
  Marquardt, and Clerk}}]{Peano2016}
\bibinfo{author}{\bibfnamefont{V.}~\bibnamefont{Peano}},
  \bibinfo{author}{\bibfnamefont{M.}~\bibnamefont{Houde}},
  \bibinfo{author}{\bibfnamefont{F.}~\bibnamefont{Marquardt}},
  \bibnamefont{and} \bibinfo{author}{\bibfnamefont{A.~A.} \bibnamefont{Clerk}},
  \emph{\bibinfo{title}{{Topological Quantum Fluctuations and Traveling Wave
  Amplifiers}}}, \bibinfo{journal}{Physical Review X}
  \textbf{\bibinfo{volume}{6}}, \bibinfo{pages}{041026} (\bibinfo{year}{2016}).

\bibitem[{\citenamefont{Gao et~al.}(2018)\citenamefont{Gao, Lester, Zhang,
  Wang, Rosenblum, Frunzio, Jiang, Girvin, and Schoelkopf}}]{Gao2018}
\bibinfo{author}{\bibfnamefont{Y.~Y.} \bibnamefont{Gao}},
  \bibinfo{author}{\bibfnamefont{B.~J.} \bibnamefont{Lester}},
  \bibinfo{author}{\bibfnamefont{Y.}~\bibnamefont{Zhang}},
  \bibinfo{author}{\bibfnamefont{C.}~\bibnamefont{Wang}},
  \bibinfo{author}{\bibfnamefont{S.}~\bibnamefont{Rosenblum}},
  \bibinfo{author}{\bibfnamefont{L.}~\bibnamefont{Frunzio}},
  \bibinfo{author}{\bibfnamefont{L.}~\bibnamefont{Jiang}},
  \bibinfo{author}{\bibfnamefont{S.~M.} \bibnamefont{Girvin}},
  \bibnamefont{and} \bibinfo{author}{\bibfnamefont{R.~J.}
  \bibnamefont{Schoelkopf}}, \emph{\bibinfo{title}{{Programmable Interference
  between Two Microwave Quantum Memories}}}, \bibinfo{journal}{Physical Review
  X} \textbf{\bibinfo{volume}{8}}, \bibinfo{pages}{021073}
  (\bibinfo{year}{2018}).

\bibitem[{\citenamefont{Gao et~al.}(2019)\citenamefont{Gao, Lester, Chou,
  Frunzio, Devoret, Jiang, Girvin, and Schoelkopf}}]{Gao2018a}
\bibinfo{author}{\bibfnamefont{Y.~Y.} \bibnamefont{Gao}},
  \bibinfo{author}{\bibfnamefont{B.~J.} \bibnamefont{Lester}},
  \bibinfo{author}{\bibfnamefont{K.~S.} \bibnamefont{Chou}},
  \bibinfo{author}{\bibfnamefont{L.}~\bibnamefont{Frunzio}},
  \bibinfo{author}{\bibfnamefont{M.~H.} \bibnamefont{Devoret}},
  \bibinfo{author}{\bibfnamefont{L.}~\bibnamefont{Jiang}},
  \bibinfo{author}{\bibfnamefont{S.~M.} \bibnamefont{Girvin}},
  \bibnamefont{and} \bibinfo{author}{\bibfnamefont{R.~J.}
  \bibnamefont{Schoelkopf}}, \emph{\bibinfo{title}{{Entanglement of bosonic
  modes through an engineered exchange interaction}}},
  \bibinfo{journal}{Nature} \textbf{\bibinfo{volume}{566}},
  \bibinfo{pages}{509} (\bibinfo{year}{2019}).

\bibitem[{\citenamefont{Gottesman et~al.}(2001)\citenamefont{Gottesman, Kitaev,
  and Preskill}}]{Gottesman2001}
\bibinfo{author}{\bibfnamefont{D.}~\bibnamefont{Gottesman}},
  \bibinfo{author}{\bibfnamefont{A.}~\bibnamefont{Kitaev}}, \bibnamefont{and}
  \bibinfo{author}{\bibfnamefont{J.}~\bibnamefont{Preskill}},
  \emph{\bibinfo{title}{{Encoding a qubit in an oscillator}}},
  \bibinfo{journal}{Physical Review A} \textbf{\bibinfo{volume}{64}},
  \bibinfo{pages}{012310} (\bibinfo{year}{2001}).

\bibitem[{\citenamefont{Terhal and Weigand}(2016)}]{Terhal2016}
\bibinfo{author}{\bibfnamefont{B.~M.} \bibnamefont{Terhal}} \bibnamefont{and}
  \bibinfo{author}{\bibfnamefont{D.}~\bibnamefont{Weigand}},
  \emph{\bibinfo{title}{{Encoding a qubit into a cavity mode in circuit QED
  using phase estimation}}}, \bibinfo{journal}{Physical Review A}
  \textbf{\bibinfo{volume}{93}}, \bibinfo{pages}{012315}
  (\bibinfo{year}{2016}).

\bibitem[{\citenamefont{Eichler and Wallraff}(2014)}]{Eichler2013}
\bibinfo{author}{\bibfnamefont{C.}~\bibnamefont{Eichler}} \bibnamefont{and}
  \bibinfo{author}{\bibfnamefont{A.}~\bibnamefont{Wallraff}},
  \emph{\bibinfo{title}{{Controlling the dynamic range of a Josephson
  parametric amplifier}}}, \bibinfo{journal}{EPJ Quantum Technology}
  \textbf{\bibinfo{volume}{1}}, \bibinfo{pages}{2} (\bibinfo{year}{2014}).

\bibitem[{\citenamefont{Vlastakis et~al.}(2013)\citenamefont{Vlastakis,
  Kirchmair, Leghtas, Nigg, Frunzio, Girvin, Mirrahimi, Devoret, and
  Schoelkopf}}]{Vlastakis2013}
\bibinfo{author}{\bibfnamefont{B.}~\bibnamefont{Vlastakis}},
  \bibinfo{author}{\bibfnamefont{G.}~\bibnamefont{Kirchmair}},
  \bibinfo{author}{\bibfnamefont{Z.}~\bibnamefont{Leghtas}},
  \bibinfo{author}{\bibfnamefont{S.~E.} \bibnamefont{Nigg}},
  \bibinfo{author}{\bibfnamefont{L.}~\bibnamefont{Frunzio}},
  \bibinfo{author}{\bibfnamefont{S.~M.} \bibnamefont{Girvin}},
  \bibinfo{author}{\bibfnamefont{M.}~\bibnamefont{Mirrahimi}},
  \bibinfo{author}{\bibfnamefont{M.~H.} \bibnamefont{Devoret}},
  \bibnamefont{and} \bibinfo{author}{\bibfnamefont{R.~J.}
  \bibnamefont{Schoelkopf}}, \emph{\bibinfo{title}{{Deterministically Encoding
  Quantum Information Using 100-Photon Schrodinger Cat States}}},
  \bibinfo{journal}{Science} \textbf{\bibinfo{volume}{342}},
  \bibinfo{pages}{607} (\bibinfo{year}{2013}).

\bibitem[{\citenamefont{Kirchmair et~al.}(2013)\citenamefont{Kirchmair,
  Vlastakis, Leghtas, Nigg, Paik, Ginossar, Mirrahimi, Frunzio, Girvin, and
  Schoelkopf}}]{Kirchmair2013}
\bibinfo{author}{\bibfnamefont{G.}~\bibnamefont{Kirchmair}},
  \bibinfo{author}{\bibfnamefont{B.}~\bibnamefont{Vlastakis}},
  \bibinfo{author}{\bibfnamefont{Z.}~\bibnamefont{Leghtas}},
  \bibinfo{author}{\bibfnamefont{S.~E.} \bibnamefont{Nigg}},
  \bibinfo{author}{\bibfnamefont{H.}~\bibnamefont{Paik}},
  \bibinfo{author}{\bibfnamefont{E.}~\bibnamefont{Ginossar}},
  \bibinfo{author}{\bibfnamefont{M.}~\bibnamefont{Mirrahimi}},
  \bibinfo{author}{\bibfnamefont{L.}~\bibnamefont{Frunzio}},
  \bibinfo{author}{\bibfnamefont{S.~M.} \bibnamefont{Girvin}},
  \bibnamefont{and} \bibinfo{author}{\bibfnamefont{R.~J.}
  \bibnamefont{Schoelkopf}}, \emph{\bibinfo{title}{{Observation of quantum
  state collapse and revival due to the single-photon Kerr effect}}},
  \bibinfo{journal}{Nature} \textbf{\bibinfo{volume}{495}},
  \bibinfo{pages}{205} (\bibinfo{year}{2013}).

\bibitem[{\citenamefont{Boutin et~al.}(2017)\citenamefont{Boutin, Toyli,
  Venkatramani, Eddins, Siddiqi, and Blais}}]{Boutin2017}
\bibinfo{author}{\bibfnamefont{S.}~\bibnamefont{Boutin}},
  \bibinfo{author}{\bibfnamefont{D.~M.} \bibnamefont{Toyli}},
  \bibinfo{author}{\bibfnamefont{A.~V.} \bibnamefont{Venkatramani}},
  \bibinfo{author}{\bibfnamefont{A.~W.} \bibnamefont{Eddins}},
  \bibinfo{author}{\bibfnamefont{I.}~\bibnamefont{Siddiqi}}, \bibnamefont{and}
  \bibinfo{author}{\bibfnamefont{A.}~\bibnamefont{Blais}},
  \emph{\bibinfo{title}{{Effect of Higher-Order Nonlinearities on Amplification
  and Squeezing in Josephson Parametric Amplifiers}}},
  \bibinfo{journal}{Physical Review Applied} \textbf{\bibinfo{volume}{8}},
  \bibinfo{pages}{054030} (\bibinfo{year}{2017}).

\bibitem[{\citenamefont{Malnou et~al.}(2018)\citenamefont{Malnou, Palken, Vale,
  Hilton, and Lehnert}}]{Malnou2018}
\bibinfo{author}{\bibfnamefont{M.}~\bibnamefont{Malnou}},
  \bibinfo{author}{\bibfnamefont{D.~A.} \bibnamefont{Palken}},
  \bibinfo{author}{\bibfnamefont{L.~R.} \bibnamefont{Vale}},
  \bibinfo{author}{\bibfnamefont{G.~C.} \bibnamefont{Hilton}},
  \bibnamefont{and} \bibinfo{author}{\bibfnamefont{K.~W.}
  \bibnamefont{Lehnert}}, \emph{\bibinfo{title}{{Optimal Operation of a
  Josephson Parametric Amplifier for Vacuum Squeezing}}},
  \bibinfo{journal}{Physical Review Applied} \textbf{\bibinfo{volume}{9}},
  \bibinfo{pages}{044023} (\bibinfo{year}{2018}).

\bibitem[{\citenamefont{Liu et~al.}(2017)\citenamefont{Liu, Chien, Cao, Lanes,
  Alpern, Pekker, and Hatridge}}]{Liu2017}
\bibinfo{author}{\bibfnamefont{G.}~\bibnamefont{Liu}},
  \bibinfo{author}{\bibfnamefont{T.-C.} \bibnamefont{Chien}},
  \bibinfo{author}{\bibfnamefont{X.}~\bibnamefont{Cao}},
  \bibinfo{author}{\bibfnamefont{O.}~\bibnamefont{Lanes}},
  \bibinfo{author}{\bibfnamefont{E.}~\bibnamefont{Alpern}},
  \bibinfo{author}{\bibfnamefont{D.}~\bibnamefont{Pekker}}, \bibnamefont{and}
  \bibinfo{author}{\bibfnamefont{M.}~\bibnamefont{Hatridge}},
  \emph{\bibinfo{title}{{Josephson parametric converter saturation and higher
  order effects}}}, \bibinfo{journal}{Applied Physics Letters}
  \textbf{\bibinfo{volume}{111}}, \bibinfo{pages}{202603}
  (\bibinfo{year}{2017}).

\bibitem[{\citenamefont{Frattini et~al.}(2018)\citenamefont{Frattini, Sivak,
  Lingenfelter, Shankar, and Devoret}}]{Frattini2018}
\bibinfo{author}{\bibfnamefont{N.~E.} \bibnamefont{Frattini}},
  \bibinfo{author}{\bibfnamefont{V.~V.} \bibnamefont{Sivak}},
  \bibinfo{author}{\bibfnamefont{A.}~\bibnamefont{Lingenfelter}},
  \bibinfo{author}{\bibfnamefont{S.}~\bibnamefont{Shankar}}, \bibnamefont{and}
  \bibinfo{author}{\bibfnamefont{M.~H.} \bibnamefont{Devoret}},
  \emph{\bibinfo{title}{{Optimizing the Nonlinearity and Dissipation of a SNAIL
  Parametric Amplifier for Dynamic Range}}}, \bibinfo{journal}{Physical Review
  Applied} \textbf{\bibinfo{volume}{10}}, \bibinfo{pages}{054020}
  (\bibinfo{year}{2018}).

\bibitem[{\citenamefont{Planat et~al.}(2019)\citenamefont{Planat,
  Dassonneville, Mart{\'{i}}nez, Foroughi, Buisson, Hasch-Guichard, Naud,
  Vijay, Murch, and Roch}}]{Planat2018}
\bibinfo{author}{\bibfnamefont{L.}~\bibnamefont{Planat}},
  \bibinfo{author}{\bibfnamefont{R.}~\bibnamefont{Dassonneville}},
  \bibinfo{author}{\bibfnamefont{J.~P.} \bibnamefont{Mart{\'{i}}nez}},
  \bibinfo{author}{\bibfnamefont{F.}~\bibnamefont{Foroughi}},
  \bibinfo{author}{\bibfnamefont{O.}~\bibnamefont{Buisson}},
  \bibinfo{author}{\bibfnamefont{W.}~\bibnamefont{Hasch-Guichard}},
  \bibinfo{author}{\bibfnamefont{C.}~\bibnamefont{Naud}},
  \bibinfo{author}{\bibfnamefont{R.}~\bibnamefont{Vijay}},
  \bibinfo{author}{\bibfnamefont{K.}~\bibnamefont{Murch}}, \bibnamefont{and}
  \bibinfo{author}{\bibfnamefont{N.}~\bibnamefont{Roch}},
  \emph{\bibinfo{title}{{Understanding the Saturation Power of Josephson
  Parametric Amplifiers Made from SQUID Arrays}}}, \bibinfo{journal}{Physical
  Review Applied} \textbf{\bibinfo{volume}{11}}, \bibinfo{pages}{034014}
  (\bibinfo{year}{2019}).

\bibitem[{\citenamefont{O'Brien et~al.}(2014)\citenamefont{O'Brien, Macklin,
  Siddiqi, and Zhang}}]{OBrien2014}
\bibinfo{author}{\bibfnamefont{K.}~\bibnamefont{O'Brien}},
  \bibinfo{author}{\bibfnamefont{C.}~\bibnamefont{Macklin}},
  \bibinfo{author}{\bibfnamefont{I.}~\bibnamefont{Siddiqi}}, \bibnamefont{and}
  \bibinfo{author}{\bibfnamefont{X.}~\bibnamefont{Zhang}},
  \emph{\bibinfo{title}{{Resonant Phase Matching of Josephson Junction
  Traveling Wave Parametric Amplifiers}}}, \bibinfo{journal}{Physical Review
  Letters} \textbf{\bibinfo{volume}{113}}, \bibinfo{pages}{157001}
  (\bibinfo{year}{2014}).

\bibitem[{\citenamefont{Macklin et~al.}(2015)\citenamefont{Macklin, O'Brien,
  Hover, Schwartz, Bolkhovsky, Zhang, Oliver, and Siddiqi}}]{Macklin2015}
\bibinfo{author}{\bibfnamefont{C.}~\bibnamefont{Macklin}},
  \bibinfo{author}{\bibfnamefont{K.}~\bibnamefont{O'Brien}},
  \bibinfo{author}{\bibfnamefont{D.}~\bibnamefont{Hover}},
  \bibinfo{author}{\bibfnamefont{M.~E.} \bibnamefont{Schwartz}},
  \bibinfo{author}{\bibfnamefont{V.}~\bibnamefont{Bolkhovsky}},
  \bibinfo{author}{\bibfnamefont{X.}~\bibnamefont{Zhang}},
  \bibinfo{author}{\bibfnamefont{W.~D.} \bibnamefont{Oliver}},
  \bibnamefont{and} \bibinfo{author}{\bibfnamefont{I.}~\bibnamefont{Siddiqi}},
  \emph{\bibinfo{title}{{A near-quantum-limited Josephson traveling-wave
  parametric amplifier}}}, \bibinfo{journal}{Science}
  \textbf{\bibinfo{volume}{350}}, \bibinfo{pages}{307} (\bibinfo{year}{2015}).

\bibitem[{\citenamefont{Vissers et~al.}(2016)\citenamefont{Vissers, Erickson,
  Ku, Vale, Wu, Hilton, and Pappas}}]{Vissers2015}
\bibinfo{author}{\bibfnamefont{M.~R.} \bibnamefont{Vissers}},
  \bibinfo{author}{\bibfnamefont{R.~P.} \bibnamefont{Erickson}},
  \bibinfo{author}{\bibfnamefont{H.-S.} \bibnamefont{Ku}},
  \bibinfo{author}{\bibfnamefont{L.}~\bibnamefont{Vale}},
  \bibinfo{author}{\bibfnamefont{X.}~\bibnamefont{Wu}},
  \bibinfo{author}{\bibfnamefont{G.~C.} \bibnamefont{Hilton}},
  \bibnamefont{and} \bibinfo{author}{\bibfnamefont{D.~P.}
  \bibnamefont{Pappas}}, \emph{\bibinfo{title}{{Low-noise kinetic inductance
  traveling-wave amplifier using three-wave mixing}}},
  \bibinfo{journal}{Applied Physics Letters} \textbf{\bibinfo{volume}{108}},
  \bibinfo{pages}{012601} (\bibinfo{year}{2016}).

\bibitem[{\citenamefont{Ranzani et~al.}(2018)\citenamefont{Ranzani, Bal, Fong,
  Ribeill, Wu, Long, Ku, Erickson, Pappas, and Ohki}}]{Ranzani2018}
\bibinfo{author}{\bibfnamefont{L.}~\bibnamefont{Ranzani}},
  \bibinfo{author}{\bibfnamefont{M.}~\bibnamefont{Bal}},
  \bibinfo{author}{\bibfnamefont{K.~C.} \bibnamefont{Fong}},
  \bibinfo{author}{\bibfnamefont{G.}~\bibnamefont{Ribeill}},
  \bibinfo{author}{\bibfnamefont{X.}~\bibnamefont{Wu}},
  \bibinfo{author}{\bibfnamefont{J.}~\bibnamefont{Long}},
  \bibinfo{author}{\bibfnamefont{H.-S.} \bibnamefont{Ku}},
  \bibinfo{author}{\bibfnamefont{R.~P.} \bibnamefont{Erickson}},
  \bibinfo{author}{\bibfnamefont{D.}~\bibnamefont{Pappas}}, \bibnamefont{and}
  \bibinfo{author}{\bibfnamefont{T.~A.} \bibnamefont{Ohki}},
  \emph{\bibinfo{title}{{Kinetic inductance traveling-wave amplifiers for
  multiplexed qubit readout}}}, \bibinfo{journal}{Applied Physics Letters}
  \textbf{\bibinfo{volume}{113}}, \bibinfo{pages}{242602}
  (\bibinfo{year}{2018}).

\bibitem[{\citenamefont{Yamamoto et~al.}(2008)\citenamefont{Yamamoto, Inomata,
  Watanabe, Matsuba, Miyazaki, Oliver, Nakamura, and Tsai}}]{Yamamoto2008}
\bibinfo{author}{\bibfnamefont{T.}~\bibnamefont{Yamamoto}},
  \bibinfo{author}{\bibfnamefont{K.}~\bibnamefont{Inomata}},
  \bibinfo{author}{\bibfnamefont{M.}~\bibnamefont{Watanabe}},
  \bibinfo{author}{\bibfnamefont{K.}~\bibnamefont{Matsuba}},
  \bibinfo{author}{\bibfnamefont{T.}~\bibnamefont{Miyazaki}},
  \bibinfo{author}{\bibfnamefont{W.~D.} \bibnamefont{Oliver}},
  \bibinfo{author}{\bibfnamefont{Y.}~\bibnamefont{Nakamura}}, \bibnamefont{and}
  \bibinfo{author}{\bibfnamefont{J.~S.} \bibnamefont{Tsai}},
  \emph{\bibinfo{title}{{Flux-driven Josephson parametric amplifier}}},
  \bibinfo{journal}{Applied Physics Letters} \textbf{\bibinfo{volume}{93}},
  \bibinfo{pages}{042510} (\bibinfo{year}{2008}).

\bibitem[{\citenamefont{Zhou et~al.}(2014)\citenamefont{Zhou, Schmitt, Bertet,
  Vion, Wustmann, Shumeiko, and Esteve}}]{Zhou2014}
\bibinfo{author}{\bibfnamefont{X.}~\bibnamefont{Zhou}},
  \bibinfo{author}{\bibfnamefont{V.}~\bibnamefont{Schmitt}},
  \bibinfo{author}{\bibfnamefont{P.}~\bibnamefont{Bertet}},
  \bibinfo{author}{\bibfnamefont{D.}~\bibnamefont{Vion}},
  \bibinfo{author}{\bibfnamefont{W.}~\bibnamefont{Wustmann}},
  \bibinfo{author}{\bibfnamefont{V.}~\bibnamefont{Shumeiko}}, \bibnamefont{and}
  \bibinfo{author}{\bibfnamefont{D.}~\bibnamefont{Esteve}},
  \emph{\bibinfo{title}{{High-gain weakly nonlinear flux-modulated Josephson
  parametric amplifier using a SQUID array}}}, \bibinfo{journal}{Physical
  Review B} \textbf{\bibinfo{volume}{89}}, \bibinfo{pages}{214517}
  (\bibinfo{year}{2014}).

\bibitem[{\citenamefont{Simoen et~al.}(2015)\citenamefont{Simoen, Chang,
  Krantz, Bylander, Wustmann, Shumeiko, Delsing, and Wilson}}]{Simoen2015}
\bibinfo{author}{\bibfnamefont{M.}~\bibnamefont{Simoen}},
  \bibinfo{author}{\bibfnamefont{C.~W.~S.} \bibnamefont{Chang}},
  \bibinfo{author}{\bibfnamefont{P.}~\bibnamefont{Krantz}},
  \bibinfo{author}{\bibfnamefont{J.}~\bibnamefont{Bylander}},
  \bibinfo{author}{\bibfnamefont{W.}~\bibnamefont{Wustmann}},
  \bibinfo{author}{\bibfnamefont{V.}~\bibnamefont{Shumeiko}},
  \bibinfo{author}{\bibfnamefont{P.}~\bibnamefont{Delsing}}, \bibnamefont{and}
  \bibinfo{author}{\bibfnamefont{C.~M.} \bibnamefont{Wilson}},
  \emph{\bibinfo{title}{{Characterization of a multimode coplanar waveguide
  parametric amplifier}}}, \bibinfo{journal}{Journal of Applied Physics}
  \textbf{\bibinfo{volume}{118}}, \bibinfo{pages}{154501}
  (\bibinfo{year}{2015}).

\bibitem[{\citenamefont{Abdo et~al.}(2013)\citenamefont{Abdo, Kamal, and
  Devoret}}]{Abdo2013}
\bibinfo{author}{\bibfnamefont{B.}~\bibnamefont{Abdo}},
  \bibinfo{author}{\bibfnamefont{A.}~\bibnamefont{Kamal}}, \bibnamefont{and}
  \bibinfo{author}{\bibfnamefont{M.}~\bibnamefont{Devoret}},
  \emph{\bibinfo{title}{{Nondegenerate three-wave mixing with the Josephson
  ring modulator}}}, \bibinfo{journal}{Physical Review B}
  \textbf{\bibinfo{volume}{87}}, \bibinfo{pages}{014508}
  (\bibinfo{year}{2013}).

\bibitem[{\citenamefont{Bergeal et~al.}(2010)\citenamefont{Bergeal, Vijay,
  Manucharyan, Siddiqi, Schoelkopf, Girvin, and Devoret}}]{Bergeal2010}
\bibinfo{author}{\bibfnamefont{N.}~\bibnamefont{Bergeal}},
  \bibinfo{author}{\bibfnamefont{R.}~\bibnamefont{Vijay}},
  \bibinfo{author}{\bibfnamefont{V.~E.} \bibnamefont{Manucharyan}},
  \bibinfo{author}{\bibfnamefont{I.}~\bibnamefont{Siddiqi}},
  \bibinfo{author}{\bibfnamefont{R.~J.} \bibnamefont{Schoelkopf}},
  \bibinfo{author}{\bibfnamefont{S.~M.} \bibnamefont{Girvin}},
  \bibnamefont{and} \bibinfo{author}{\bibfnamefont{M.~H.}
  \bibnamefont{Devoret}}, \emph{\bibinfo{title}{{Analog information processing
  at the quantum limit with a Josephson ring modulator}}},
  \bibinfo{journal}{Nature Physics} \textbf{\bibinfo{volume}{6}},
  \bibinfo{pages}{296} (\bibinfo{year}{2010}).

\bibitem[{\citenamefont{Frattini et~al.}(2017)\citenamefont{Frattini, Vool,
  Shankar, Narla, Sliwa, and Devoret}}]{Frattini2017}
\bibinfo{author}{\bibfnamefont{N.~E.} \bibnamefont{Frattini}},
  \bibinfo{author}{\bibfnamefont{U.}~\bibnamefont{Vool}},
  \bibinfo{author}{\bibfnamefont{S.}~\bibnamefont{Shankar}},
  \bibinfo{author}{\bibfnamefont{A.}~\bibnamefont{Narla}},
  \bibinfo{author}{\bibfnamefont{K.~M.} \bibnamefont{Sliwa}}, \bibnamefont{and}
  \bibinfo{author}{\bibfnamefont{M.~H.} \bibnamefont{Devoret}},
  \emph{\bibinfo{title}{{3-Wave Mixing Josephson Dipole Element}}},
  \bibinfo{journal}{Applied Physics Letters} \textbf{\bibinfo{volume}{110}},
  \bibinfo{pages}{222603} (\bibinfo{year}{2017}).

\bibitem[{\citenamefont{Yurke et~al.}(1989)\citenamefont{Yurke, Corruccini,
  Kaminsky, Rupp, Smith, Silver, Simon, and Whittaker}}]{Yurke1989}
\bibinfo{author}{\bibfnamefont{B.}~\bibnamefont{Yurke}},
  \bibinfo{author}{\bibfnamefont{L.~R.} \bibnamefont{Corruccini}},
  \bibinfo{author}{\bibfnamefont{P.~G.} \bibnamefont{Kaminsky}},
  \bibinfo{author}{\bibfnamefont{L.~W.} \bibnamefont{Rupp}},
  \bibinfo{author}{\bibfnamefont{A.~D.} \bibnamefont{Smith}},
  \bibinfo{author}{\bibfnamefont{A.~H.} \bibnamefont{Silver}},
  \bibinfo{author}{\bibfnamefont{R.~W.} \bibnamefont{Simon}}, \bibnamefont{and}
  \bibinfo{author}{\bibfnamefont{E.~A.} \bibnamefont{Whittaker}},
  \emph{\bibinfo{title}{{Observation of parametric amplification and
  deamplification in a Josephson parametric amplifier}}},
  \bibinfo{journal}{Physical Review A} \textbf{\bibinfo{volume}{39}},
  \bibinfo{pages}{2519} (\bibinfo{year}{1989}).

\bibitem[{\citenamefont{Zorin}(2016)}]{Zorin2016}
\bibinfo{author}{\bibfnamefont{A.~B.} \bibnamefont{Zorin}},
  \emph{\bibinfo{title}{{Josephson traveling-wave parametric amplifier with
  three-wave mixing}}}, \bibinfo{journal}{Physical Review Applied}
  \textbf{\bibinfo{volume}{6}}, \bibinfo{pages}{034006} (\bibinfo{year}{2016}).

\bibitem[{\citenamefont{Chien et~al.}(2019)\citenamefont{Chien, Lanes, Liu,
  Cao, Lu, Motz, Liu, Pekker, and Hatridge}}]{Chien2019}
\bibinfo{author}{\bibfnamefont{T.~C.} \bibnamefont{Chien}},
  \bibinfo{author}{\bibfnamefont{O.}~\bibnamefont{Lanes}},
  \bibinfo{author}{\bibfnamefont{C.}~\bibnamefont{Liu}},
  \bibinfo{author}{\bibfnamefont{X.}~\bibnamefont{Cao}},
  \bibinfo{author}{\bibfnamefont{P.}~\bibnamefont{Lu}},
  \bibinfo{author}{\bibfnamefont{S.}~\bibnamefont{Motz}},
  \bibinfo{author}{\bibfnamefont{G.}~\bibnamefont{Liu}},
  \bibinfo{author}{\bibfnamefont{D.}~\bibnamefont{Pekker}}, \bibnamefont{and}
  \bibinfo{author}{\bibfnamefont{M.}~\bibnamefont{Hatridge}},
  \emph{\bibinfo{title}{{Multiparametric Amplification and Qubit Measurement
  with a Kerr-free Josephson Ring Modulator}}} (\bibinfo{year}{2019}),
  \eprint{arXiv:1903.02102}.

\bibitem[{\citenamefont{Sundqvist and Delsing}(2014)}]{Sundqvist2014}
\bibinfo{author}{\bibfnamefont{K.~M.} \bibnamefont{Sundqvist}}
  \bibnamefont{and} \bibinfo{author}{\bibfnamefont{P.}~\bibnamefont{Delsing}},
  \emph{\bibinfo{title}{{Negative-resistance models for parametrically
  flux-pumped superconducting quantum interference devices}}},
  \bibinfo{journal}{EPJ Quantum Technology} \textbf{\bibinfo{volume}{1}},
  \bibinfo{pages}{6} (\bibinfo{year}{2014}).

\bibitem[{\citenamefont{Verney et~al.}(2019)\citenamefont{Verney, Lescanne,
  Devoret, Leghtas, and Mirrahimi}}]{Verney2018}
\bibinfo{author}{\bibfnamefont{L.}~\bibnamefont{Verney}},
  \bibinfo{author}{\bibfnamefont{R.}~\bibnamefont{Lescanne}},
  \bibinfo{author}{\bibfnamefont{M.~H.} \bibnamefont{Devoret}},
  \bibinfo{author}{\bibfnamefont{Z.}~\bibnamefont{Leghtas}}, \bibnamefont{and}
  \bibinfo{author}{\bibfnamefont{M.}~\bibnamefont{Mirrahimi}},
  \emph{\bibinfo{title}{{Structural Instability of Driven Josephson Circuits
  Prevented by an Inductive Shunt}}}, \bibinfo{journal}{Physical Review
  Applied} \textbf{\bibinfo{volume}{11}}, \bibinfo{pages}{024003}
  (\bibinfo{year}{2019}).

\bibitem[{\citenamefont{Kochetov and Fedorov}(2015)}]{Kochetov2015}
\bibinfo{author}{\bibfnamefont{B.~A.} \bibnamefont{Kochetov}} \bibnamefont{and}
  \bibinfo{author}{\bibfnamefont{A.}~\bibnamefont{Fedorov}},
  \emph{\bibinfo{title}{{Higher-order nonlinear effects in a Josephson
  parametric amplifier}}}, \bibinfo{journal}{Physical Review B - Condensed
  Matter and Materials Physics} \textbf{\bibinfo{volume}{92}},
  \bibinfo{pages}{224304} (\bibinfo{year}{2015}).

\bibitem[{\citenamefont{Dykman et~al.}(1998)\citenamefont{Dykman, Maloney,
  Smelyanskiy, and Silverstein}}]{Dykman1998}
\bibinfo{author}{\bibfnamefont{M.~I.} \bibnamefont{Dykman}},
  \bibinfo{author}{\bibfnamefont{C.~M.} \bibnamefont{Maloney}},
  \bibinfo{author}{\bibfnamefont{V.~N.} \bibnamefont{Smelyanskiy}},
  \bibnamefont{and}
  \bibinfo{author}{\bibfnamefont{M.}~\bibnamefont{Silverstein}},
  \emph{\bibinfo{title}{{Fluctuational phase-flip transitions in parametrically
  driven oscillators}}}, \bibinfo{journal}{Physical Review E}
  \textbf{\bibinfo{volume}{57}}, \bibinfo{pages}{5202} (\bibinfo{year}{1998}).

\bibitem[{\citenamefont{Wustmann and Shumeiko}(2013)}]{Wustmann2013}
\bibinfo{author}{\bibfnamefont{W.}~\bibnamefont{Wustmann}} \bibnamefont{and}
  \bibinfo{author}{\bibfnamefont{V.}~\bibnamefont{Shumeiko}},
  \emph{\bibinfo{title}{{Parametric resonance in tunable superconducting
  cavities}}}, \bibinfo{journal}{Physical Review B}
  \textbf{\bibinfo{volume}{87}}, \bibinfo{pages}{184501}
  (\bibinfo{year}{2013}).

\bibitem[{\citenamefont{Zorin and Makhlin}(2011)}]{Zorin2011}
\bibinfo{author}{\bibfnamefont{A.~B.} \bibnamefont{Zorin}} \bibnamefont{and}
  \bibinfo{author}{\bibfnamefont{Y.}~\bibnamefont{Makhlin}},
  \emph{\bibinfo{title}{{Period-doubling bifurcation readout for a Josephson
  qubit}}}, \bibinfo{journal}{Physical Review B} \textbf{\bibinfo{volume}{83}},
  \bibinfo{pages}{224506} (\bibinfo{year}{2011}).

\bibitem[{\citenamefont{Lin et~al.}(2014)\citenamefont{Lin, Inomata, Koshino,
  Oliver, Nakamura, Tsai, and Yamamoto}}]{Lin2014}
\bibinfo{author}{\bibfnamefont{Z.}~\bibnamefont{Lin}},
  \bibinfo{author}{\bibfnamefont{K.}~\bibnamefont{Inomata}},
  \bibinfo{author}{\bibfnamefont{K.}~\bibnamefont{Koshino}},
  \bibinfo{author}{\bibfnamefont{W.}~\bibnamefont{Oliver}},
  \bibinfo{author}{\bibfnamefont{Y.}~\bibnamefont{Nakamura}},
  \bibinfo{author}{\bibfnamefont{J.}~\bibnamefont{Tsai}}, \bibnamefont{and}
  \bibinfo{author}{\bibfnamefont{T.}~\bibnamefont{Yamamoto}},
  \emph{\bibinfo{title}{{Josephson parametric phase-locked oscillator and its
  application to dispersive readout of superconducting qubits}}},
  \bibinfo{journal}{Nature Communications} \textbf{\bibinfo{volume}{5}},
  \bibinfo{pages}{4480} (\bibinfo{year}{2014}).

\bibitem[{\citenamefont{Dykman}(2012)}]{Dykman2012}
\bibinfo{author}{\bibfnamefont{M.}~\bibnamefont{Dykman}},
  \emph{\bibinfo{title}{{Fluctuating Nonlinear Oscillators: From Nanomechanics
  to Quantum Superconducting Circuits}}} (\bibinfo{publisher}{Oxford University
  Press}, \bibinfo{address}{Oxford}, \bibinfo{year}{2012}).

\bibitem[{\citenamefont{Krantz et~al.}(2016)\citenamefont{Krantz, Bengtsson,
  Simoen, Gustavsson, Shumeiko, Oliver, Wilson, Delsing, and
  Bylander}}]{Krantz2016}
\bibinfo{author}{\bibfnamefont{P.}~\bibnamefont{Krantz}},
  \bibinfo{author}{\bibfnamefont{A.}~\bibnamefont{Bengtsson}},
  \bibinfo{author}{\bibfnamefont{M.}~\bibnamefont{Simoen}},
  \bibinfo{author}{\bibfnamefont{S.}~\bibnamefont{Gustavsson}},
  \bibinfo{author}{\bibfnamefont{V.}~\bibnamefont{Shumeiko}},
  \bibinfo{author}{\bibfnamefont{W.~D.} \bibnamefont{Oliver}},
  \bibinfo{author}{\bibfnamefont{C.~M.} \bibnamefont{Wilson}},
  \bibinfo{author}{\bibfnamefont{P.}~\bibnamefont{Delsing}}, \bibnamefont{and}
  \bibinfo{author}{\bibfnamefont{J.}~\bibnamefont{Bylander}},
  \emph{\bibinfo{title}{{Single-shot read-out of a superconducting qubit using
  a Josephson parametric oscillator}}}, \bibinfo{journal}{Nature
  Communications} \textbf{\bibinfo{volume}{7}}, \bibinfo{pages}{11417}
  (\bibinfo{year}{2016}).

\bibitem[{\citenamefont{Svensson et~al.}(2017)\citenamefont{Svensson,
  Bengtsson, Krantz, Bylander, Shumeiko, and Delsing}}]{Svensson2017a}
\bibinfo{author}{\bibfnamefont{I.-M.} \bibnamefont{Svensson}},
  \bibinfo{author}{\bibfnamefont{A.}~\bibnamefont{Bengtsson}},
  \bibinfo{author}{\bibfnamefont{P.}~\bibnamefont{Krantz}},
  \bibinfo{author}{\bibfnamefont{J.}~\bibnamefont{Bylander}},
  \bibinfo{author}{\bibfnamefont{V.}~\bibnamefont{Shumeiko}}, \bibnamefont{and}
  \bibinfo{author}{\bibfnamefont{P.}~\bibnamefont{Delsing}},
  \emph{\bibinfo{title}{{Period-tripling subharmonic oscillations in a driven
  superconducting resonator}}}, \bibinfo{journal}{Physical Review B}
  \textbf{\bibinfo{volume}{96}}, \bibinfo{pages}{174503}
  (\bibinfo{year}{2017}).

\bibitem[{\citenamefont{Lin et~al.}(2015)\citenamefont{Lin, Nakamura, and
  Dykman}}]{Lin2015}
\bibinfo{author}{\bibfnamefont{Z.~R.} \bibnamefont{Lin}},
  \bibinfo{author}{\bibfnamefont{Y.}~\bibnamefont{Nakamura}}, \bibnamefont{and}
  \bibinfo{author}{\bibfnamefont{M.~I.} \bibnamefont{Dykman}},
  \emph{\bibinfo{title}{{Critical fluctuations and the rates of interstate
  switching near the excitation threshold of a quantum parametric
  oscillator}}}, \bibinfo{journal}{Physical Review E}
  \textbf{\bibinfo{volume}{92}}, \bibinfo{pages}{022105}
  (\bibinfo{year}{2015}).

\bibitem[{\citenamefont{Zhang et~al.}(2017)\citenamefont{Zhang, Huang,
  Gershenson, and Bell}}]{Zhang2017}
\bibinfo{author}{\bibfnamefont{W.}~\bibnamefont{Zhang}},
  \bibinfo{author}{\bibfnamefont{W.}~\bibnamefont{Huang}},
  \bibinfo{author}{\bibfnamefont{M.~E.} \bibnamefont{Gershenson}},
  \bibnamefont{and} \bibinfo{author}{\bibfnamefont{M.~T.} \bibnamefont{Bell}},
  \emph{\bibinfo{title}{{Josephson Metamaterial with a Widely Tunable Positive
  or Negative Kerr Constant}}}, \bibinfo{journal}{Physical Review Applied}
  \textbf{\bibinfo{volume}{8}}, \bibinfo{pages}{051001} (\bibinfo{year}{2017}).

\bibitem[{\citenamefont{Mutus et~al.}(2014)\citenamefont{Mutus, White, Barends,
  Chen, Chen, Chiaro, Dunsworth, Jeffrey, Kelly, Megrant et~al.}}]{Mutus2014}
\bibinfo{author}{\bibfnamefont{J.~Y.} \bibnamefont{Mutus}},
  \bibinfo{author}{\bibfnamefont{T.~C.} \bibnamefont{White}},
  \bibinfo{author}{\bibfnamefont{R.}~\bibnamefont{Barends}},
  \bibinfo{author}{\bibfnamefont{Y.}~\bibnamefont{Chen}},
  \bibinfo{author}{\bibfnamefont{Z.}~\bibnamefont{Chen}},
  \bibinfo{author}{\bibfnamefont{B.}~\bibnamefont{Chiaro}},
  \bibinfo{author}{\bibfnamefont{A.}~\bibnamefont{Dunsworth}},
  \bibinfo{author}{\bibfnamefont{E.}~\bibnamefont{Jeffrey}},
  \bibinfo{author}{\bibfnamefont{J.}~\bibnamefont{Kelly}},
  \bibinfo{author}{\bibfnamefont{A.}~\bibnamefont{Megrant}},
  \bibnamefont{et~al.}, \emph{\bibinfo{title}{{Strong environmental coupling in
  a Josephson parametric amplifier}}}, \bibinfo{journal}{Applied Physics
  Letters} \textbf{\bibinfo{volume}{104}}, \bibinfo{pages}{263513}
  (\bibinfo{year}{2014}).

\bibitem[{\citenamefont{Roy et~al.}(2015)\citenamefont{Roy, Kundu, Chand,
  Vadiraj, Ranadive, Nehra, Patankar, Aumentado, Clerk, and Vijay}}]{Roy2015}
\bibinfo{author}{\bibfnamefont{T.}~\bibnamefont{Roy}},
  \bibinfo{author}{\bibfnamefont{S.}~\bibnamefont{Kundu}},
  \bibinfo{author}{\bibfnamefont{M.}~\bibnamefont{Chand}},
  \bibinfo{author}{\bibfnamefont{A.~M.} \bibnamefont{Vadiraj}},
  \bibinfo{author}{\bibfnamefont{A.}~\bibnamefont{Ranadive}},
  \bibinfo{author}{\bibfnamefont{N.}~\bibnamefont{Nehra}},
  \bibinfo{author}{\bibfnamefont{M.~P.} \bibnamefont{Patankar}},
  \bibinfo{author}{\bibfnamefont{J.}~\bibnamefont{Aumentado}},
  \bibinfo{author}{\bibfnamefont{A.~A.} \bibnamefont{Clerk}}, \bibnamefont{and}
  \bibinfo{author}{\bibfnamefont{R.}~\bibnamefont{Vijay}},
  \emph{\bibinfo{title}{{Broadband parametric amplification with impedance
  engineering: Beyond the gain-bandwidth product}}}, \bibinfo{journal}{Applied
  Physics Letters} \textbf{\bibinfo{volume}{107}}, \bibinfo{pages}{262601}
  (\bibinfo{year}{2015}).

\bibitem[{\citenamefont{Spietz et~al.}(2003)\citenamefont{Spietz, Lehnert,
  Siddiqi, and Schoelkopf}}]{Spietz2003}
\bibinfo{author}{\bibfnamefont{L.}~\bibnamefont{Spietz}},
  \bibinfo{author}{\bibfnamefont{K.}~\bibnamefont{Lehnert}},
  \bibinfo{author}{\bibfnamefont{I.}~\bibnamefont{Siddiqi}}, \bibnamefont{and}
  \bibinfo{author}{\bibfnamefont{R.~J.} \bibnamefont{Schoelkopf}},
  \emph{\bibinfo{title}{{Primary Electronic Thermometry Using the Shot Noise of
  a Tunnel Junction}}}, \bibinfo{journal}{Science}
  \textbf{\bibinfo{volume}{300}}, \bibinfo{pages}{1929} (\bibinfo{year}{2003}).

\bibitem[{\citenamefont{Spietz et~al.}(2010)\citenamefont{Spietz, Irwin, Lee,
  and Aumentado}}]{Spietz2010}
\bibinfo{author}{\bibfnamefont{L.}~\bibnamefont{Spietz}},
  \bibinfo{author}{\bibfnamefont{K.}~\bibnamefont{Irwin}},
  \bibinfo{author}{\bibfnamefont{M.}~\bibnamefont{Lee}}, \bibnamefont{and}
  \bibinfo{author}{\bibfnamefont{J.}~\bibnamefont{Aumentado}},
  \emph{\bibinfo{title}{{Noise performance of lumped element direct current
  superconducting quantum interference device amplifiers in the 4–8 GHz
  range}}}, \bibinfo{journal}{Applied Physics Letters}
  \textbf{\bibinfo{volume}{97}}, \bibinfo{pages}{142502}
  (\bibinfo{year}{2010}).

\bibitem[{\citenamefont{Touzard et~al.}(2019)\citenamefont{Touzard, Kou,
  Frattini, Sivak, Puri, Grimm, Frunzio, Shankar, and Devoret}}]{Touzard2018}
\bibinfo{author}{\bibfnamefont{S.}~\bibnamefont{Touzard}},
  \bibinfo{author}{\bibfnamefont{A.}~\bibnamefont{Kou}},
  \bibinfo{author}{\bibfnamefont{N.~E.} \bibnamefont{Frattini}},
  \bibinfo{author}{\bibfnamefont{V.~V.} \bibnamefont{Sivak}},
  \bibinfo{author}{\bibfnamefont{S.}~\bibnamefont{Puri}},
  \bibinfo{author}{\bibfnamefont{A.}~\bibnamefont{Grimm}},
  \bibinfo{author}{\bibfnamefont{L.}~\bibnamefont{Frunzio}},
  \bibinfo{author}{\bibfnamefont{S.}~\bibnamefont{Shankar}}, \bibnamefont{and}
  \bibinfo{author}{\bibfnamefont{M.~H.} \bibnamefont{Devoret}},
  \emph{\bibinfo{title}{{Gated Conditional Displacement Readout of
  Superconducting Qubits}}}, \bibinfo{journal}{Physical Review Letters}
  \textbf{\bibinfo{volume}{122}}, \bibinfo{pages}{080502}
  (\bibinfo{year}{2019}).

\bibitem[{\citenamefont{Landau and Lifshitz}(1969)}]{Landau}
\bibinfo{author}{\bibfnamefont{L.~D.} \bibnamefont{Landau}} \bibnamefont{and}
  \bibinfo{author}{\bibfnamefont{E.~M.} \bibnamefont{Lifshitz}},
  \emph{\bibinfo{title}{{Statistical Physics}}} (\bibinfo{publisher}{Pergamon},
  \bibinfo{address}{Oxford}, \bibinfo{year}{1969}).

\bibitem[{\citenamefont{Abdo et~al.}(2017)\citenamefont{Abdo, Chavez-Garcia,
  Brink, Keefe, and Chow}}]{Abdo2017}
\bibinfo{author}{\bibfnamefont{B.}~\bibnamefont{Abdo}},
  \bibinfo{author}{\bibfnamefont{J.~M.} \bibnamefont{Chavez-Garcia}},
  \bibinfo{author}{\bibfnamefont{M.}~\bibnamefont{Brink}},
  \bibinfo{author}{\bibfnamefont{G.}~\bibnamefont{Keefe}}, \bibnamefont{and}
  \bibinfo{author}{\bibfnamefont{J.~M.} \bibnamefont{Chow}},
  \emph{\bibinfo{title}{{Time-multiplexed amplification in a hybrid-less and
  coil-less Josephson parametric converter}}}, \bibinfo{journal}{Applied
  Physics Letters} \textbf{\bibinfo{volume}{110}}, \bibinfo{pages}{082601}
  (\bibinfo{year}{2017}).

\bibitem[{\citenamefont{Gardiner and Zoller}(2004)}]{Gardiner2004}
\bibinfo{author}{\bibfnamefont{C.}~\bibnamefont{Gardiner}} \bibnamefont{and}
  \bibinfo{author}{\bibfnamefont{P.}~\bibnamefont{Zoller}},
  \emph{\bibinfo{title}{{Quantum Noise}}}
  (\bibinfo{publisher}{Springer-Verlag}, \bibinfo{address}{Heidelberg},
  \bibinfo{year}{2004}).

\bibitem[{\citenamefont{Naaman et~al.}(2017)\citenamefont{Naaman, Ferguson, and
  Epstein}}]{Naaman2017}
\bibinfo{author}{\bibfnamefont{O.}~\bibnamefont{Naaman}},
  \bibinfo{author}{\bibfnamefont{D.~G.} \bibnamefont{Ferguson}},
  \bibnamefont{and} \bibinfo{author}{\bibfnamefont{R.~J.}
  \bibnamefont{Epstein}}, \emph{\bibinfo{title}{{High Saturation Power
  Josephson Parametric Amplifier with GHz Bandwidth}}} (\bibinfo{year}{2017}),
  \eprint{arXiv:1711.07549}.

\bibitem[{\citenamefont{Manucharyan}(2012)}]{Manucharyan2012}
\bibinfo{author}{\bibfnamefont{V.}~\bibnamefont{Manucharyan}}, Ph.D. thesis,
  \bibinfo{school}{Yale University} (\bibinfo{year}{2012}).

\bibitem[{\citenamefont{Wallquist et~al.}(2006)\citenamefont{Wallquist,
  Shumeiko, and Wendin}}]{Wallquist2006}
\bibinfo{author}{\bibfnamefont{M.}~\bibnamefont{Wallquist}},
  \bibinfo{author}{\bibfnamefont{V.~S.} \bibnamefont{Shumeiko}},
  \bibnamefont{and} \bibinfo{author}{\bibfnamefont{G.}~\bibnamefont{Wendin}},
  \emph{\bibinfo{title}{{Selective coupling of superconducting charge qubits
  mediated by a tunable stripline cavity}}}, \bibinfo{journal}{Physical Review
  B} \textbf{\bibinfo{volume}{74}}, \bibinfo{pages}{224506}
  (\bibinfo{year}{2006}).

\bibitem[{\citenamefont{Nigg et~al.}(2012)\citenamefont{Nigg, Paik, Vlastakis,
  Kirchmair, Shankar, Frunzio, Devoret, Schoelkopf, and Girvin}}]{Nigg2012}
\bibinfo{author}{\bibfnamefont{S.~E.} \bibnamefont{Nigg}},
  \bibinfo{author}{\bibfnamefont{H.}~\bibnamefont{Paik}},
  \bibinfo{author}{\bibfnamefont{B.}~\bibnamefont{Vlastakis}},
  \bibinfo{author}{\bibfnamefont{G.}~\bibnamefont{Kirchmair}},
  \bibinfo{author}{\bibfnamefont{S.}~\bibnamefont{Shankar}},
  \bibinfo{author}{\bibfnamefont{L.}~\bibnamefont{Frunzio}},
  \bibinfo{author}{\bibfnamefont{M.~H.} \bibnamefont{Devoret}},
  \bibinfo{author}{\bibfnamefont{R.~J.} \bibnamefont{Schoelkopf}},
  \bibnamefont{and} \bibinfo{author}{\bibfnamefont{S.~M.}
  \bibnamefont{Girvin}}, \emph{\bibinfo{title}{{Black-Box Superconducting
  Circuit Quantization}}}, \bibinfo{journal}{Physical Review Letters}
  \textbf{\bibinfo{volume}{108}}, \bibinfo{pages}{240502}
  (\bibinfo{year}{2012}).

\end{thebibliography}

\end{document}